\documentclass[fleqn,useAMS,usenatbib]{mnras}

\usepackage{enumerate}
\usepackage{amsmath,amssymb,amsfonts}
\usepackage{epsf,lscape}
\usepackage{color}
\usepackage{multicol}        % Multi-column entries in tables
\usepackage[dvipdfmx]{graphicx}

\usepackage[T1]{fontenc}
\usepackage{ae,aecompl}

\title[BOSS-CMASS SHAM I]{Connecting massive galaxies to dark matter halos in BOSS - I. 
Is galaxy color a stochastic process in high-mass halos?
}

\author[S. Saito et al.]{
\parbox{\textwidth}{Shun~Saito$^{1}$\thanks{E-mail: shun.saito@ipmu.jp}, 
Alexie~Leauthaud$^{1}$,
Andrew~P.~Hearin$^{2}$,
Kevin~Bundy$^{1}$,
Andrew R. Zentner$^{3}$,
Peter S. Behroozi$^{4,5}$, 
Beth A. Reid$^{4,6}$, 
Manodeep Sinha$^{7}$, 
Jean Coupon$^{8}$, 
Jeremy L. Tinker$^{9}$, 
Martin White$^{6,10,11}$, 
Donald P. Schneider$^{12,13}$}
\\
\\
$^{1}$Kavli Institute for the Physics and Mathematics of the Universe (WPI), 
The University of Tokyo Institutes for Advanced Study,\\
The University of Tokyo, Chiba 277-8582, Japan\\
$^{2}$Yale Center for Astronomy and Astrophysics, Yale University, New Haven, CT 06511, USA\\
$^{3}$Department of Physics and Astronomy and Pittsburgh Particle physics, 
Astrophysics and Cosmology Center (PITT PACC),\\
University of Pittsburgh, Pittsburgh, PA 15260\\
$^{4}$Hubble fellow\\
$^{5}$Space Telescope Science Institute, Baltimore, MD21218, USA\\
$^{6}$Lawrence Berkeley National Laboratory, 1 Cyclotron Road, Berkeley, CA 94720, USA\\
$^{7}$Department of Physics and Astronomy, Vanderbilt University, Nashville, TN, 37235\\
$^{8}$Astronomical Observatory of the University of Geneva, ch. d’Ecogia 16, 1290 Versoix, Switzerland\\
$^{9}$Center for Cosmology and Particle Physics, Department of Physics, New York University, New York, NY 10003, USA\\
$^{10}$Department of Physics, 366 LeConte Hall, University of California at Berkeley, Berkeley, CA 94720, USA\\
$^{11}$Department of Astronomy, 601 Campbell Hall, University of California at Berkeley, Berkeley, CA 94720, USA\\
$^{12}$Department of Astronomy and Astrophysics, The Pennsylvania State University, University Park, PA 16802, USA\\
$^{13}$Institute for Gravitation and the Cosmos, The Pennsylvania State University, University Park, PA 16802, USA
}

\begin{document}

\label{firstpage}
\maketitle
%----------------------------------------------

\begin{abstract}
  We use subhalo abundance matching (SHAM) to model the stellar mass
  function (SMF) and clustering of the Baryon Oscillation
  Spectroscopic Survey (BOSS) ``CMASS'' sample at $z\sim0.5$. We
  introduce a novel method which accounts for the stellar mass
  incompleteness of CMASS as a function of redshift, and produce CMASS
  mock catalogs which include selection effects, reproduce the overall
  SMF, the projected two-point correlation function $w_{\rm p}$, the
  CMASS $dn/dz$, and are made publicly available. We study the effects
  of assembly bias above collapse mass in the context of ``age
  matching'' and show that these effects are markedly different
  compared to the ones explored by Hearin et al. (2013) at lower
  stellar masses.  We construct two models, one in which galaxy color
  is stochastic (``AbM'' model) as well as a model which contains
  assembly bias effects (``AgM'' model). By confronting the redshift
  dependent clustering of CMASS with the predictions from our model,
  we argue that that galaxy colors are not a stochastic process in
  high-mass halos. Our results suggest that the colors of galaxies in high-mass
  halos are determined by other halo properties besides halo peak
  velocity and that assembly bias effects play an important role
  in determining the clustering properties of this sample.
\end{abstract}


%%%%%%%%%%%%%%%%%%%%%%%%%%%%%%%%%%%%%%%%%%%%%%%%%%%%%%%%%%%%%%%%%%%%%%%%%%%%
%%%%%%%%%%%%%%%%%%%%%%%%%%%%%%%%%%%%%%%%%%%%%%%%%%%%%%%%%%%%%%%%%%%%%%%%%%%%
\section{Introduction}
%%%%%%%%%%%%%%%%%%%%%%%%%%%%%%%%%%%%%%%%%%%%%%%%%%%%%%%%%%%%%%%%%%%%%%%%%%%%
%%%%%%%%%%%%%%%%%%%%%%%%%%%%%%%%%%%%%%%%%%%%%%%%%%%%%%%%%%%%%%%%%%%%%%%%%%%%

The overall picture that galaxies form, evolve, and reside in dark
matter halos that assemble hierarchically has gained consensus by
passing a variety of observational tests over a wide range of cosmic
history \citep[for a review, see][]{Mo_2010book}.  However,
understanding the detailed relation between galaxies and dark matter
halos is critical in order to form a more concrete theory of galaxy
formation and evolution. In particular, unveiling how the stellar
masses and star-formation properties of galaxies depend on halo
properties is still a topic of active investigation.  For low-mass
galaxies ($M_*\lesssim 10^{11}M_{\odot}$), recent studies of galaxy
clustering and galaxy-galaxy lensing suggest that red and blue
galaxies live in halos of different mass at fixed stellar mass at
$0\lesssim z \lesssim 1$
\citep[][]{Zehavi_2005,Mandelbaum_2006,Tinker_2013,Coupon_2015, Mandelbaum:2015lr} or
that at fixed stellar mass, galaxy color may correlate with halo age
\citep[][]{Hearin_2014}.

While many previous studies focus on low or intermediate mass galaxies, 
the galaxy-halo mass connection is also worth investigating 
for the most massive galaxies in the universe. 
The majority of galaxies with masses $M_*\gtrsim 10^{11}M_{\odot}$ are 
centrals hosted by massive halos ($M_{\rm halo}\gtrsim 10^{13}M_{\odot}$) 
\citep[][]{White_2011,Leauthaud_2011,Coupon_2015}. 
From a theoretical standpoint, gas in these high-mass halos is thought to 
be heated by pressure-supported shocks \citep[the so-called ``hot halo mode",][]{Dekel_2006}. 
In addition, at these halo masses, ``maintenance mode" feed-back mechanisms 
such as radio-mode feedback are thought to further limit star-formation 
in the most massive galaxies \citep[e.g.,][]{Croton_2006}. Observationally, 
however, not all massive galaxies are systematically ``red and dead". 
For example, although they are rare, brightest cluster galaxies associated 
with cool core clusters can exhibit star formation rates of order 
$\mathcal{O}(10\mathchar`- 100)\,M_{\odot}{\rm yr^{-1}}$ 
(e.g., in Abell 1835 at $z\sim 0.25$ and in Perseus A and Cygnus A at $z\sim 0.1$) 
\citep[e.g.,][]{Liu_2012,McDonald_2012,Fraser_McKelvie_2014}. 
At group scales, \citet{Tinker_2012} found that as many as 20\% of 
central galaxies in halos with $\log_{10}(M_{\rm halo}/M_{\odot})>13$ at $z\sim 0.5$ 
have sufficient levels of star formation to exhibit blue colors. 
A key question is then: what determines color in high mass halos? 
Is star formation in massive galaxies simply a stochastic process 
due to episodic amount of gas cooling and/or due to mergers with gas rich satellites? 
Or are the colors of massive galaxies more fundamentally linked to assembly history 
of their parent dark matter halos?

Large spectroscopic samples of massive galaxies are of tremendous
value in addressing these types of questions. Spectroscopic redshifts
are crucial for computing precise measurements of galaxy-clustering
and galaxy-galaxy lensing which can be used to constrain the
galaxy-halo connection
\citep[e.g.,][]{Mandelbaum_2006,Leauthaud_2011,Coupon_2015}.  The
availability of spectroscopic redshifts also reduces uncertainties on
stellar mass estimates. Spectroscopic surveys such as zCOSMOS
\citep[][]{Lilly_2007}, VVDS \citep[][]{Le_F_vre_2015}, DEEP2
\citep[][]{Newman_2013}, PRIMUS \citep[][]{Coil_2011}, and VIPERS
\citep[][]{Guzzo_2014}, however, cover relatively small areas ranging
from a few square degrees to a few tens of square degrees and do not
provide statistically significant samples of the most massive galaxies
($\log_{10}(M_*/M_{\odot})>11.5$). For this reason, we turn our
attention instead to the Sloan Digital Sky Survey III
\citep[SDSS-III,][]{Eisenstein_2011} Baryon Oscillation Spectroscopic
Survey \citep[BOSS,][]{Dawson_2013}.  The main BOSS cosmological
sample, the so-called {\it CMASS} sample \citep[][]{Reid:2016kq}, 
includes roughly half a million massive galaxies at $\log_{10}(M_*/M_{\odot})\gtrsim 11.0$ 
at $0.43<z<0.70$ and covers a gigantic volume of approximately
$2.5\,(h^{-1}{\rm Gpc})^{3}$ at the tenth data release (DR10)
\citep[][]{Ahn_2014}. This gigantic dataset enables high signal-to-noise 
ratio measurements of three dimensional galaxy clustering on
large scales (typically separation of $r \gtrsim 10\,{\rm Mpc}$) and
provides the most accurate measurement of the Baryon Acoustic
Oscillation (BAO) scale and the Redshift-Space Distortion (RSD) signal
with a precision in DR11 \citep[][]{SDSSIII_DR12_2015} of $\approx
1\%$ and $\approx 10\%$ respectively
\citep[e.g.,][]{Anderson_2014,Beutler_2014a,Beutler_2014b,Samushia_2014}.

The main goal of this paper is to model the connection between galaxy
mass, color, and halo mass for massive galaxies using the BOSS CMASS
dataset. In addition to providing insight on the evolution of massive
galaxies, a detailed understanding of the CMASS-halo connection is
also critical because BOSS analysis pipelines need to be
systematically tested against {\it realistic} CMASS mock
catalogs. Mock catalogs within the BOSS collaboration \citep[e.g.,][]{White_2011,Manera_2012,White_2013,Kitaura_2013} 
are typically based on the Halo Occupation Distribution (HOD) approach
\citep[see e.g.,][]{Berlind_2002,Zheng_2005}. However, until present,
most studies have assumed that CMASS is a homogeneous sample and have
ignored any redshift-dependent selection effects.

%This method makes assumptions for
%the functional forms of central and satellite occupation functions
%with free parameters that are determined by fitting to $w_p$, for
%example
%\citep[e.g.,][]{White_2011,Manera_2012,White_2013,Kitaura_2013}. The
%HOD approach has also been used to model the three-dimensional
%correlation function in redshift space for the full CMASS sample
%\citep[hereafter R14]{Reid_2014}.  

Indeed, one difficulty with the CMASS sample that affects both studies
of massive galaxies as well as the creation of realistic mock
catalogs, is accounting for the selection function of the sample. 
The CMASS selection algorithm was roughly designed to select a ``{\it constant stellar-mass}" sample 
and is often quoted as being mass
limited at $\log_{10}(M_{*}/M_{\odot})> 11.3$ over the redshift range
$0.43<z<0.7$.  However, \citet{Leauthaud_2015} (hereafter, L15)
demonstrate that CMASS is only 80\% complete at
$\log_{10}(M_*/M_{\odot})>11.6$ in the narrow redshift range
$0.51<z<0.61$.  Our paper improves on previous analyses by presenting
a careful treatment of the stellar mass completeness of the CMASS
sample in our models. 

To model the CMASS-halo connection we adopt the subhalo abundance
matching (SHAM) technique. SHAM is a fairly simple and empirical
approach which assumes that galaxy properties such as luminosity or
stellar mass are monotonically related to (sub)halo properties such as
mass or circular velocity \citep[see
e.g.,][]{Kravtsov_2004,Vale_2004,Conroy_2006,Moster_2010,Behroozi_2010}.
Although there are model ambiguities in this method (e.g., in choosing
which properties to relate and how scatter is introduced), SHAM
requires relatively few parameters and also provides a straightforward
prescription for linking galaxy properties to dark matter halos in
numerical $N$-body simulations.  Our work can be considered as an
update to \citet{Nuza_2013} who used the SHAM approach to model the
CMASS-halo connection but without accounting for the stellar mass
completeness of the CMASS sample.

%This feature also enables $w_{p}$ to
%be computed directly from $N$-body simulations without resorting to
%analytic approximations which rely on fitting functions for the halo
%mass function and scale dependent bias that are uncertain at the 5\%
%level \citep[e.g.,][]{Tinker_2007,Tinker_2008,Tinker_2010}. 

In addition to the standard implementation of SHAM, we also explore
the {\it age matching} model introduced by \citet{Hearin_2013}
(hereafter, H13) which introduces galaxy color by assuming that at
fixed stellar mass, redder galaxies reside in older sub-halos. The age
matching scheme can qualitatively explain a variety of observed
statistics in the SDSS main galaxy sample including color-dependent
galaxy clustering \citep[][]{Hearin_2013,Watson_2014}, magnitude gap
statistics in galaxy groups \citep[][]{Hearin_2013b}, galaxy-galaxy
lensing \citep[][]{Hearin_2014}, galaxy conformity
\citep[][]{Hearin_2014c}, and halo mass dependence of the specific
star formation rate \citep[][]{Lim_2015}.

%This
%ansatz is motivated by the fact that the clustering of dark matter
%halos in $N$-body simulations depends on additional parameters beyond
%halo mass such as halo formation epoch, an effect known as {\it
%  assembly bias}
%\citep[e.g.,][]{Gao_2005,Wechsler_2006,Jing_2007,Gao_2007,Dalal_2008,Li_2008,Mao_2015}.

Our models are constrained by three observables: the clustering of
CMASS on radial scales $r \lesssim 0.1\mathchar`- 10\,{\rm Mpc}$, the
galaxy stellar mass function (SMF), and the SMF of CMASS galaxies as a
function of redshift.

This paper is organized as follows. The observational data are
summarized in Section \ref{sec:observation}.  Our measurements of the
correlation function and the galaxy stellar mass function are
described in Section \ref{sec:meas_stat}.  In particular, Section
\ref{sec:CMASS_SMFz} presents our measurements of the
redshift-dependent CMASS SMFs that are an essential ingredient in this
study.  Section \ref{sec:simulation} briefly summarizes the simulated
subhalo catalog.  Section \ref{sec:methodology} is a detailed
presentation of our SHAM and age matching methodology.  Our results
are described in Section \ref{sec:Results} and discussed in Section
\ref{sec:Discussion}.  Finally we summarize and conclude our study in
Section \ref{sec:summary}.

Our measurements assume a flat $\Lambda$CDM cosmology with $\Omega_{\rm m}=0.274$ and $H_{0}=70\,{\rm km\,s^{-1}\,Mpc^{-1}}$. 
For all quantities related to $w_{\rm p}$, or to $N$-body simulations, we adopt  $H_{0}=100h\,{\rm km\,s^{-1}\,Mpc^{-1}}$ and hence distance 
and mass units are written as $h^{-1}{\rm Mpc}$ and $h^{-1}{\rm M_{\odot}}$. 
Note that there are small differences between this choice and the cosmological parameters assumed for the $N$-body simulations 
that we introduce in Section \ref{sec:simulation}.

%%%%%%%%%%%%%%%%%%%%%%%%%%%%%%%%%%%%%%%%%%%%%%%%%%%%%%%%%%%%%%%%%%%%%%%%%%%%
%%%%%%%%%%%%%%%%%%%%%%%%%%%%%%%%%%%%%%%%%%%%%%%%%%%%%%%%%%%%%%%%%%%%%%%%%%%%
\section{Observational Data}
\label{sec:observation}
%%%%%%%%%%%%%%%%%%%%%%%%%%%%%%%%%%%%%%%%%%%%%%%%%%%%%%%%%%%%%%%%%%%%%%%%%%%%
%%%%%%%%%%%%%%%%%%%%%%%%%%%%%%%%%%%%%%%%%%%%%%%%%%%%%%%%%%%%%%%%%%%%%%%%%%%%
This section begins with a brief review of the BOSS DR10 CMASS sample. 
In addition to the BOSS sample, our analysis also relies on data 
from the SDSS Stripe 82 region which is roughly two magnitudes deeper than 
the SDSS main survey.  

%%%%%%%%%%%%%%%%%%%%%%%%%%%%%%%%%%%%%%%%%%%%%%%%%%%%%%%%%%%%%%%%%%%%%%%%%%%%
\subsection{The BOSS DR10 CMASS sample}
%%%%%%%%%%%%%%%%%%%%%%%%%%%%%%%%%%%%%%%%%%%%%%%%%%%%%%%%%%%%%%%%%%%%%%%%%%%%
The BOSS survey \citep[][]{Dawson_2013} is a part of SDSS-III which measured 
1.5 million spectroscopic redshifts of luminous galaxies and 160,000 quasars 
over an extragalactic footprint covering $\sim 10000\,{\rm deg}^{2}$.  
Spectroscopic observations were obtained using the 1000 object fiber-fed 
BOSS spectrograph \citep[][]{Smee_2013} on the 2.5 m aperture Sloan Foundation 
Telescope at the Apache Point Observatory \citep[][]{Gunn_1998,Gunn_2006}. 
The BOSS pipeline is described in \citet{Bolton_2012}, and 
BOSS galaxies were selected from Data Release 8
\citep[DR8,][]{Aihara_2011} {\it ugriz} photometry \citep[][]{Fukugita_1996}. 
The main purpose of BOSS is to measure the BAO 
feature and RSD from galaxy clustering. 
The internal data release 11 (DR11) and the final DR12 dataset are 
made public in \citet{SDSSIII_DR12_2015}, although the DR12 
large-scale structure CMASS catalog is not yet publicly available at this point. 
Using DR11 which contains nearly one million spectroscopic redshifts of 
galaxies over $\sim 8,500\,{\rm deg}^{2}$, the BOSS collaboration has measured 
BAO and RSD signals to an unprecedented precision of $1\%$ and $10\%$, 
respectively \citep[e.g.,][]{Anderson_2014,Beutler_2014a,Beutler_2014b,Samushia_2014}

The BOSS target selection is divided into two samples, a low-redshift sample 
(``LOWZ'') that selects luminous red galaxies at $z<0.43$ 
\citep[for details see][]{Tojeiro_2014} and a high-redshift sample (``CMASS'') 
that targets galaxies at $0.43<z<0.7$ \citep[][]{Reid:2016kq}. This paper focuses only on 
the CMASS sample which is selected using a series of color-magnitude cuts 
motivated by stellar population models from \citet{Maraston_2009}.  
The CMASS sample is selected as:
%%%%%
\begin{eqnarray}
17.5 < i_{\rm cmod} < 19.9, \nonumber\\
r_{\rm mod}-i_{\rm mod} < 2.0, \nonumber\\
d_{\perp} > 0.55, \nonumber\\
i_{\rm fib2} < 21.5, \nonumber\\
i_{\rm cmod} < 19.86 + 1.6(d_{\perp} {-} 0.8),
\label{eq:CMASS_selection}
\end{eqnarray}
%%%%%
where 
%%%%%
\begin{equation}
d_{\perp} = r_{\rm mod} {-} i_{\rm mod} {-} (g_{\rm mod} {-} r_{\rm mod})/8.0. 
\end{equation}
%%%%%
Model magnitudes are denoted with the subscript `${\rm mod}$', 
composite model magnitudes are denoted with 
the subscript `${\rm cmod}$', 
fiber magnitude within a $2''$ aperture is denoted with 
the subscript  `${\rm fib2}$'. The BOSS color cuts are
computed using model magnitudes, whereas magnitude cuts 
are computed using cmodel magnitudes. All magnitudes are 
corrected for Galactic extinction using the dust maps of 
\citet{Schlegel_1998}.\par

In this paper, we use the CMASS sample from the public DR10 dataset 
\citep[][]{Ahn_2014} that includes 409,365 galaxies over $4,892\,{\rm deg}^{2}$ 
in the North Galactic Cap (NGC) and 112,593 galaxies over $1,432\,{\rm deg}^{2}$ 
in the South Galactic Cap (SGC). Note that these numbers differ from 
those reported in \citet{Anderson_2014} simply because we adopt a different 
weighting scheme for our clustering measurements (see following section). 
While previous studies have focused on sub-samples of CMASS in limited redshift 
or magnitude ranges \citep[e.g.,][]{Guo_2013,Miyatake:2013lr,Guo_2014,More:2014qy}, 
in this paper we model the full CMASS sample over the full redshift range 
$0.43<z<0.7$.

%%%%%%%%%%%%%%%%%%%%%%%%%%%%%%%%%%%%%%%%%%%%%%%%%%%%%%%%%%%%%%%%%%%%%%%%%%%%%%
\subsection{Stripe 82 Co-add Catalog Combined with UKIDDS Photometry 
For Improved Stellar Mass Estimates}
\label{subsec:S82}
%%%%%%%%%%%%%%%%%%%%%%%%%%%%%%%%%%%%%%%%%%%%%%%%%%%%%%%%%%%%%%%%%%%%%%%%%%%%%%

A key aspect of our approach is the use of Stripe 82 --- a deeper but
narrower subset of the survey area --- for which it is possible to
construct a galaxy sample with a well-understood completeness
function.  Stripe 82 provides two key advantages. First, it was the
subject of repeat imaging campaigns in SDSS and therefore reaches
$ugriz$ depths that are roughly two magnitudes deeper than the
single-epoch SDSS imaging that was used to construct the BOSS target
catalog.  This added depth is critical for obtaining reliable
photometric redshifts (photo-z's) for massive galaxies
($\log_{10} (M_*/M_{\odot}) > 11$) that can be used to supplement the
color-selected BOSS samples out to $z \sim 0.7$.  Second, Stripe 82
was imaged by the UKIRT Infrared Deep Sky Survey
\citep[UKIDSS,][]{lawrence07} providing near-IR photometry for robust
stellar mass estimates.

In this paper, we use the Stripe 82 Massive Galaxy catalog (hereafter, {\sc s82-mgc}). 
The {\sc s82-mgc} catalog construction, photometric matching, redshift validation, 
masking, and other details are described in \citet{Bundy:2015rf}. 
The {\sc s82-mgc} catalog contains all classified galaxies from UKIDSS-LAS frames 
with 10$\sigma$ detection limits deeper than $YJHK = [20.2, 20.2, 20.2, 20.6]$ (AB) 
\citep[][]{Oke_1983}. 
These limits are those roughly needed for 10$\sigma$ detections in these bands of 
$z \sim 0.6$ passive galaxies with $\log_{10} (M_*/M_{\odot}) > 11.2$. 
UKIDSS and BOSS masks are applied to this catalog 
which covers a total area of 139.4 deg$^2$. 

The {\sc s82-mgc} catalog contains both spectroscopic and photometric
redshifts. For each galaxy, we adopt the spectroscopic redshift when it is available 
and use the photometric redshift otherwise. L15 demonstrate that the
impact of photo-z scatter on the high mass end of the SMF is
negligible.  Stellar masses are estimated for this catalog by applying
the SED-fitting code described in \citet{bundy10} to the SDSS+UKIDSS
PSF-matched photometry.  For a prior grid of SED templates and a
Chabrier IMF \citep{chabrier03}, an $M_*$ probability distribution is
obtained by scaling the model $M/L$ ratios by the inferred luminosity
in the observed $K$-band, or $H$-band if a $K$-band magnitude is not
available. The median of this distribution is taken as the $M_*$
estimate.

%%%%%%%%%%%%%%%%%%%%%%%%%%%%%%%%%%%%%%%%%%%%%%%%%%%%%%%%%%%%%%%%%%%%%%%%%%%%
%%%%%%%%%%%%%%%%%%%%%%%%%%%%%%%%%%%%%%%%%%%%%%%%%%%%%%%%%%%%%%%%%%%%%%%%%%%%
\section{Correlation Function and Stellar Mass Function Measurements}
\label{sec:meas_stat}
%%%%%%%%%%%%%%%%%%%%%%%%%%%%%%%%%%%%%%%%%%%%%%%%%%%%%%%%%%%%%%%%%%%%%%%%%%%%
%%%%%%%%%%%%%%%%%%%%%%%%%%%%%%%%%%%%%%%%%%%%%%%%%%%%%%%%%%%%%%%%%%%%%%%%%%%%

This section summarizes our measurements of several statistics
derived from the observational data described in the previous section.
After briefly explaining the measurement of the two-point correlation
function (note that we use the measurement computed by \citet[hereafter R14]{Reid_2014}), 
we present our measurement of the CMASS SMFs as
a function of redshift.

%%%%%%%%%%%%%%%%%%%%%%%%%%%%%%%%%%%%%%%%%%%%%%%%%%%%%%%%%%%%%%%%%%%%%%%%%%%%
\subsection{The CMASS Two-Point Correlation Function}
%%%%%%%%%%%%%%%%%%%%%%%%%%%%%%%%%%%%%%%%%%%%%%%%%%%%%%%%%%%%%%%%%%%%%%%%%%%%
In this paper we adopt the DR10 projected two-point correlation function, $w_p$, 
and the monopole and quadrupole of the correlation function, $\hat{\xi}_{\ell}$, 
and the associated covariance matrices determined by R14. 
We only give a brief summary of how these measurements were performed; 
we refer the reader to R14 for additional details. 
The two-dimensional redshift-space correlation function $\xi({\bf s})$ 
is measured using the Landy-Szalay estimator \cite{Landy_1993}:
%%%%%
\begin{eqnarray}
 \xi({\bf s}) = 
 \frac{DD(\Delta{\bf s})-2DR(\Delta{\bf s})+RR(\Delta{\bf s})}{RR(\Delta{\bf s})},
\end{eqnarray}
%%%%%
where $DD$, $DR$, and $RR$ are the data-data, data-random, and 
random-random pairs in a given bin 
$[{\bf s}-\Delta{\bf s}/2,{\bf s}+\Delta{\bf s}/2]$. 
The randoms account for the survey geometry and for the completeness factor which
depends on angular position and a radial selection function, $dn/dz$.  
The correlation function is integrated over the line-of-sight separation 
to obtain the projected correlation function \citep[][]{Davis_1983},
%%%%%
\begin{eqnarray}
  w_{p}(r_{p}) = 2\int^{r_{\pi,{\rm max}}}_{0}\,\xi(r_{p},r_{\pi})dr_{\pi}, 
\end{eqnarray}
%%%%%
where the three-dimensional pair separation ${\bf s}$ in redshift space 
is split into a component transverse ($r_{p}$) and parallel ($r_{\pi}$) 
to the line-of-sight direction. 
The integral is performed to $r_{\pi, {\rm max}}=80\,h^{-1}{\rm Mpc}$ and 
$w_p$ is measured from 0.194 $h^{-1}{\rm Mpc}$ to 25.98 $h^{-1}{\rm Mpc}$ with 
18 equally spaced logarithmic bins.  
The advantage of using the projected correlation function is that it is 
less sensitive than $\xi({\bf s})$ to the effects of galaxy peculiar velocities. 
Note that, however, we do account for the RSD effect \citep{van-den-Bosch:2013fk} 
in our modeling through the velocity of subhalos. 
The projected two-point correlation function is measured separately for the 
North and South Galactic Caps and these measurements are combined using a simple average, 
weighted by the number of CMASS galaxies in each hemisphere. \par

The $w_p$ measurement from R14 does not use the optimal weights 
(the so-called ``FKP'' weights), or the systematic weights \citep[][]{Anderson_2014}. 
The systematic weights affect large scales and hence are not relevant 
for our small-scale measurement. Also, this approach enables a fairer
comparison with our measurement of the galaxy SMF 
which does not use any weighting schemes. Weights are applied, however, 
to account for redshift failures and for fiber collisions. 
Fiber collisions are particularly important for small scale clustering 
measurements with BOSS -- the fiber-collision scale in BOSS is $62''$ 
which corresponds to a comoving scale of $\sim 0.45h^{-1}{\rm Mpc}$ at $z\sim 0.57$. 
To complicate matters, the BOSS tiling strategy also introduces 
a correlation between fiber collisions and the density field. 
R14 studied the impact of fiber collisions for the CMASS sample 
using tiled mock catalogs. They adopt a radial dependent correction scheme 
in which an angular up-weighting method is used at $r_{p}<1.09\,h^{-1}{\rm Mpc}$ 
and a nearest neighbor (NN) weighting scheme is used at larger scales. 
Finally, the correlation function is debiased for residual fiber-collision 
effects using the tiled mock catalogs. \par

%This effect can be corrected by the so-called angular 
%up-weighting method in which the $DD$ pairs in the Landy-Szalay estimator is 
%up-weighted according to the ratio of the angular correlation function of 
%the spectroscopic samples to that of the targets. However, the angular 
%up-weighting method could break down at large scales, since a fiber-collided 
%galaxy could sit in a more massive halo which have a larger large-scale bias. 

The covariance matrix for $w_{p}$, ${\bf C}_{w_{\rm p},{\rm boot}}$, is derived from 
5,000,000 realizations drawn from 200 bootstrap regions which are roughly 
equal in size and shape. An additional 10\% uncertainty due to the angular 
up-weighting method and the debiasing procedure are propagated into 
the diagonal element of the covariance matrix.  
As a result, the measurement error on $w_p$ increases 
below $r_{p}=1.09\,h^{-1}{\rm Mpc}$. Finally, the inverse covariance matrix 
is corrected following \citet{Hartlap_2007}. With $n_{\rm boot}=200$ 
and $n_{\rm bin}=18$, this leads to a 0.904 correction to 
the final inverse covariance matrix, 
${\bf C}^{-1}_{w_{\rm p},{\rm meas}}=0.904{\bf C}^{-1}_{w_{\rm p},{\rm boot}}$. \par 

In addition to $w_{\rm p}$, we will also use the monopole and quadrupole 
of the correlation function which contain information about the peculiar 
velocities of galaxies. Again, following R14, we adopt the pseudo 
multipole correlation function defined by
%%%%%
\begin{equation}
\hat{\xi}_{\ell}(s)=(2\ell+1)\int^{\mu_{\rm max}(s)}_{0}\,
d\mu\,\xi(s,\mu){\mathcal L}_{\ell}(\mu),
\end{equation}
%%%%%
where $s^{2}=r_{\rm p}^{2}+r_{\pi}^{2}$, $\mu=r_{\pi}/s$, 
and ${\mathcal L}_{\ell}(\mu)$ is the $\ell$-th order Legendre polynomial.  
The integration over the azimuthal angle $\mu$ is performed 
up to $\mu_{\rm max}(s)\equiv 0.534\,s^{-1}$ in order to minimize 
the impact of fiber collisions on the small-scale measurements. 
We refer the reader to R14 for further details.

%In order to suppress the skewed feature 
%in the inverse Wishart distribution when inverting the covariance matrix, 
%we follow \citet{Hartlap_2007} and apply the correction factor
%%%%%
%\begin{eqnarray}
%  {\bf C}^{-1}_{\rm meas} = 
%  \frac{n_{\rm boot}-n_{\rm bin}-2}{n_{\rm boot}-1}{\bf C}^{-1}_{\rm boot}
%\end{eqnarray}
%%%%%
%where $n_{\rm boot}=200$ and $n_{\rm bin}=18$ in this case, resulting in 
%the prefactor of 0.904. For related discussion we refer readers to 
%\citet{Krause_2012}, \citet{Mandelbaum_2013}, and \citet{Percival_2014}.\par
% => Too much detail here I think since a lot of this is already in Beth's paper

%%%%%%%%%%%%%%%%%%%%%%%%%%%%%%%%%%%%%%%%%%%%%%%%%%%%%%%%%%%%%%%%%%%%%%%%%%%%
\subsection{The Stripe 82 Stellar Mass Function at $z=0.55$}
\label{sec:totalSMF}
%%%%%%%%%%%%%%%%%%%%%%%%%%%%%%%%%%%%%%%%%%%%%%%%%%%%%%%%%%%%%%%%%%%%%%%%%%%%

As shown in L15, the CMASS sample is only stellar mass complete at the
high mass end and in a narrow redshift range.  To perform abundance
matching, however, we need to measure the total SMF. Indeed, for
abundance matching, a complete galaxy sample is necessary when rank
ordering galaxies versus halos.

\citet{Bundy:2015rf} present an estimate of the SMF at $z\sim0.5$ by
using the {\sc s82-mgc} catalog.  In order to compute the SMF, 
\citet{Bundy:2015rf} use a combination of spectroscopic redshifts,
supplemented with photometric redshifts (photo-z's) when a
spectroscopic redshift is not available.  We adopt a similar approach
and compute the SMF from the {\sc s82-mgc} at
$\log_{10}(M_*/M_{\odot})>10.5$ over $0.43<z<0.70$. Our analysis
assumes that the SMF does not vary over this redshift range. The
result is shown in Figure \ref{fig:TOTAL_SMF}. Error bars on the SMF
represent the square root of the diagonal component of the covariance
matrix, which is estimated from the data using 214 nearly-equal area
bootstrap regions.

Because the majority of galaxies at the high mass end have a
spectroscopic redshift, the impact of photo-z uncertainty on the
Stripe 82 SMF is negligible (see L15), i.e., the use of photometric
redshifts only adds a negligible amount of scatter in the total
stellar mass estimate and does not inflate the high mass end of the
SMF.

The left panel of Figure \ref{fig:TOTAL_SMF} presents a comparison between our SMF with
results from COSMOS \citep{Leauthaud_2011} and PRIMUS
\citep{Moustakas_2013} at similar redshifts. 
Figure \ref{fig:TOTAL_SMF} demonstrates that, because of the large area covered by Stripe 82,
the high mass end of the total SMF is tightly constrained at
$\log_{10}(M_*/M_{\odot})>11.3$ over $0.43<z<0.70$, while COSMOS and
PRIMUS constrain the low mass end. The comparison with COSMOS and
PRIMUS suggests that the {\sc s82-mgc} is complete to
$\log_{10}(M_*/M_{\odot})\sim 11.2$ at $z=0.7$ \citep[][]{Bundy:2015rf}. 

We will use the {\sc s82-mgc} SMF measured using 8 data points over
the range $11.5 \le \log_{10}(M_*/M_{\odot})\le 12.0$. The inverse
covariance matrix for the {\sc s82-mgc} SMF, ${\bf C}^{-1}_{\rm SMF}$
is computed as follows. First we compute the covariance matrix ${\bf C}_{\rm SMF,boot}$ 
from 214 bootstrap regions, and then smooth the
noise in the non-diagonal components using a boxcar algorithm
\citep{Mandelbaum_2006}.  Finally we multiply by the Hartlap
correction factor which is $\sim 0.958$, i.e., 
${\bf C}^{-1}_{\rm SMF}=0.958{\bf C}^{-1}_{\rm SMF,boot}$.  
Although the error budget is dominated by the Poisson noise which only contributes to diagonal
components \citep[][]{Smith_2012}, the Poisson error underestimate the
errors. We find that the diagonal component in our jackknife
covariance matrix is larger than the Poisson errors by a factor of
$\sim 30\%$ in the mass range of our interest.

%%%%%%%%%%%%%%%%%%%%%%%%%%%%%%%%%%%%%%%%%%%%%%%%%%%%%%%%%%%%%%%%%%%%%%%%%%%%
\subsection{SMF of CMASS galaxies as a Function of Redshift}
\label{sec:CMASS_SMFz}
%%%%%%%%%%%%%%%%%%%%%%%%%%%%%%%%%%%%%%%%%%%%%%%%%%%%%%%%%%%%%%%%%%%%%%%%%%%%

\begin{figure*}
\begin{center}
\includegraphics[width=2\columnwidth]{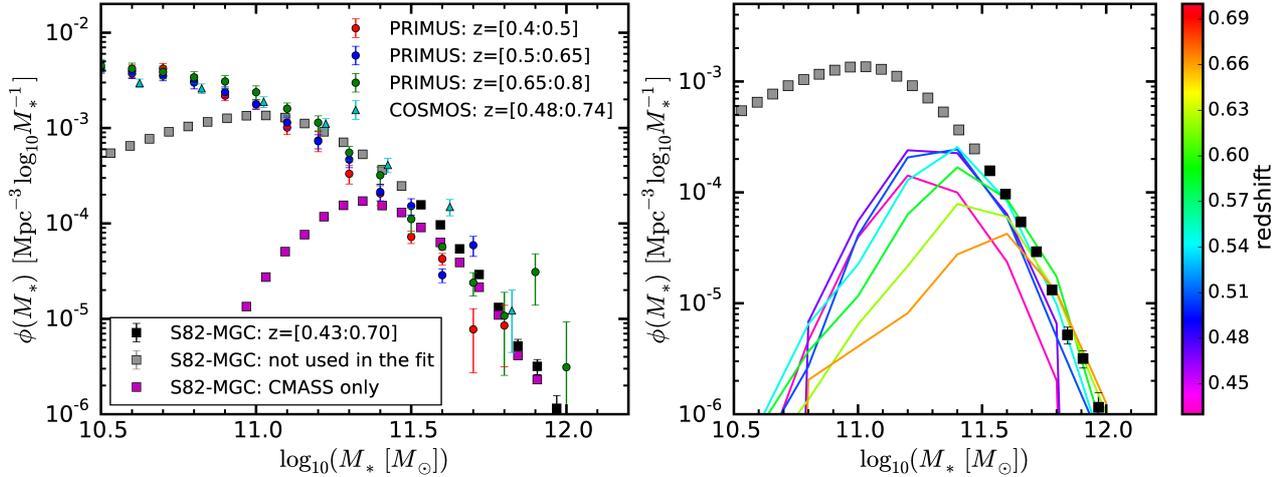}
%\begin{figure}[!ht]
%\begin{center}
%\includegraphics[width=1\columnwidth,bb=0 0 522 328]{fig/SMF_S82v5.pdf}
\caption{\label{fig:TOTAL_SMF} ({\it Left}) The total SMF from Stripe
  82 (black and grey squares) measured from {\sc s82-mcg} (139.4
  deg$^2$) and the SMF measured using only CMASS galaxies (magenta
  squares). Other SMFs determined from smaller area surveys at similar
  redshifts are also shown. Red, blue, and green circles indicate
  results from PRIMUS (5.5 deg$^2$) at $0.4<z<0.5$, $0.5<z<0.65$, and
  $0.65<z<0.8$, respectively.  Cyan triangles represent one wide
  redshift bin from the COSMOS survey (1.64 deg$^2$). 
  %The solid blackvcurve displays our best-fit to the total SMF (see Section
  %\ref{subsec:AbM}). 
  As explained later, we only use data points with $\log_{10}(M_{*}/M_{\odot})> 11.5$ 
  (black squares) when fitting against the SMF data. 
  ({\it Right})
  SMFs as a function of redshift measured using only the CMASS sample.
  As a reference, we also present the total SMF from the {\sc s82-mgc}
  at $0.43<z<0.70$ and $\log_{10}(M_{*}/M_{\odot})> 10.5$. As
  demonstrated in L15, the CMASS SMFs vary with redshift and CMASS is
  only complete in terms of stellar mass at the highest masses and in
  a relatively narrow redshift range.}
\end{center}
\end{figure*}

The other ingredient that will be important in our analysis are the
SMFs of CMASS galaxies as a function of redshift. The right panel of
Figure \ref{fig:TOTAL_SMF} shows SMFs for CMASS galaxies measured
using the {\sc s82-mgc} in 7 redshift bins with $\Delta z=0.04$. As
can be seen from the right panel of Figure \ref{fig:TOTAL_SMF}, the
completeness of CMASS depends both on redshift and stellar mass; this
is because the effects of the magnitude and color cuts that define the
CMASS sample vary with redshift. The utility of the these CMASS SMFs 
will be apparent when we describe our methodology in Section \ref{sec:methodology}. 

%To account for the stellar mass completeness of CMASS, 
%we perform abundance matching using the {\it total} SMF from Figure
%\ref{fig:TOTAL_SMF} but then down-sample our mock catalog as a
%function of stellar mass and redshift in order to simultaneously
%recover the correct $dn/dz$ as well as the correct stellar mass
%distributions for CMASS (see Section \ref{sec:methodology}).

%\begin{figure}[!ht]
%\begin{center}
%\includegraphics[width=1\columnwidth,bb=0 0 522 328]{fig/SMF_S82v5_cmass.pdf}
%\caption{\label{fig:SMF_CMASS_Z} 
%SMFs as a function of redshift
%  measured using only the CMASS sample.  As a reference, we also show
 % the total SMF from the {\sc s82-mgc} at $0.43<z<0.70$ and
 % $M_{*}\,[M_{\odot}]> 10.5\,{\rm dex}$.  Note that the  {\sc s82-mgc}
 % is complete to roughly  $M_{*}\,[M_{\odot}]> 11.2\,{\rm dex}$. As
 % shown in L15, CMASS is only complete in terms
 % of stellar mass at the highest masses and in a realtively narrow redshift
 % range.}
%\end{center}
%\end{figure}

%%%%%%%%%%%%%%%%%%%%%%%%%%%%%%%%%%%%%%%%%%%%%%%%%%%%%%%%%%%%%%%%%%%%%%%%%%%%%%%
%%%%%%%%%%%%%%%%%%%%%%%%%%%%%%%%%%%%%%%%%%%%%%%%%%%%%%%%%%%%%%%%%%%%%%%%%%%%%%%
\section{Subhalo Catalog}
\label{sec:simulation}
%%%%%%%%%%%%%%%%%%%%%%%%%%%%%%%%%%%%%%%%%%%%%%%%%%%%%%%%%%%%%%%%%%%%%%%%%%%%%%%
%%%%%%%%%%%%%%%%%%%%%%%%%%%%%%%%%%%%%%%%%%%%%%%%%%%%%%%%%%%%%%%%%%%%%%%%%%%%%%%
In this section we present the \textit{N}-body simulation and 
subhalo catalog that is an essential ingredient in 
our abundance-matching study. We also perform tests of 
the completeness of the subhalo catalog. 

%%%%%%%%%%%%%%%%%%%%%%%%%%%%%%%%%%%%%%%%%%%%%%%%%%%%%%%%%%%%%%%%%%%%%%%%%%%%%%%
\subsection{\textit{N}-body Simulation}
%%%%%%%%%%%%%%%%%%%%%%%%%%%%%%%%%%%%%%%%%%%%%%%%%%%%%%%%%%%%%%%%%%%%%%%%%%%%%%%
Because the BOSS DR10 CMASS sample covers a large comoving volume, $V
\sim 2.6\,(h^{-1}{\rm Gpc})^{3}$, with a high number density of
$\overline{n}\sim 3\times 10^{-4}\,(h^{-1}{\rm Mpc})^{-3}$, our analysis
requires a large volume {\it N}-body simulation that can resolve halos
to $10^{12}\,M_{\odot}$.  We use the publicly available MultiDark
simulation, {\sf MDR1} \citep{Prada_2012,Riebe:2013xx}. The cosmological
parameters in {\sf MDR1} are consistent with a flat WMAP5 $\Lambda$CDM
cosmology \citep{Komatsu_2009}: $\Omega_{\rm m0}=0.27$,
$\Omega_{\Lambda}=0.73$, $\Omega_{\rm b0}=0.047$, $n_{\rm s}=0.95$,
and $\sigma_{8}=0.82$. This cosmology is similar to the one used for
our measurements of the clustering signals, therefore
safely ignore the cosmological uncertainty in the distance scale
\citep{More_2013}.  {\sf MDR1} is a $L_{\rm box}=1.0\,h^{-1}{\rm Gpc}$
simulation with a particle mass of $8.7\times 10^{9}\,h^{-1}M_{\odot}$
($N_{\rm par}=2048^{3}$ particles).  We use an output at $z=0.534$
which is close to the the peak of the BOSS CMASS $dn/dz$ at 
$z\sim 0.55$. 

%%%%%%%%%%%%%%%%%%%%%%%%%%%%%%%%%%%%%%%%%%%%%%%%%%%%%%%%%%%%%%%%%%%%%%%%%%%%%%%
\subsection{Halo Catalogs and Merger Trees}
%%%%%%%%%%%%%%%%%%%%%%%%%%%%%%%%%%%%%%%%%%%%%%%%%%%%%%%%%%%%%%%%%%%%%%%%%%%%%%%
Halos and subhalos are identified using the {\sf Rockstar} algorithm
\citep[][]{Behroozi_2013a,Behroozi_2013b}. {\sf Rockstar} is a phase-space
halo finder that also considers halo merger histories to provide a
robust and stable identification of halos and subhalos.  The
advantages of {\sf Rockstar} compared to other halo finders are well
documented in \citet{Knebe_2011} and \citet{Onions_2012}. These
studies suggest that among halo finders, {\sf Rockstar} finder is the
least sensitive to resolution effects.
%especially in a region very close to the center of massive host halos 
%(e.g. $\lesssim 50\,{\rm kpc}$ for a Milky Way sized halo). 
%Note that, however, our clustering study does not require 
%such a high resolved simulation as we will see below 
%{\bf [AL: meaning unclear here. "high" compared to what?? -> SS: Added comment in a bracket above]}. 
%{\sf Rockstar} also produces halo merger trees and catalogs with 
%various parameters derived from the halo assembly history, 
%including for example, $V_{\rm peak}$, the maximum halo circular velocity.\par 
{\sf Rockstar}, with the Consistent Trees algorithm, produces halo merger trees and 
catalogs with various parameters derived from the halo assembly history, 
including, for example, $V_{\rm peak}$, the maximum halo circular velocity for each subhalo.
%This is because, while \citet{Knebe_2011} discussed $V_{\rm max}$ 
%can be a good parameter to rank order subhalos, $V_{\rm peak}$ 
%is less sensitive to effect of e.g. the tidal stripping 
%\cite{Conroy_2006,Trujillo_Gomez_2011,Reddick_2013}.\par 

%We account for the redshift dependence of the stellar-mass completeness of CMASS 
%by abundance matching {\sf MDR1} in thin redshift slices 
%from $z=0.43$ to $z=0.7$,  
%and down sample each slice to match 
%the observed CMASS $dn/dz$ (see Section \ref{subsec:mstar_incompleteness}). 

We use the ``$Z$''-axis of the {\sf MDR1} simulation 
as the line-of-sight direction. In order to maximize 
the volume of our mock, we re-map the 1-$h^{-1}$Gpc  {\sf MDR1} 
cube into a cuboid of dimensions, 
$(X,Y,Z)=(3.7417,0.4082,0.6547)L_{\rm box}$ 
following the method developed by \citet{Carlson_2010}. 
After remapping, the $Z$-axis has a length of $654.7\,h^{-1}{\rm Mpc}$ 
corresponding to a redshift range of $0.42<z<0.71$. 
This includes a margin that is sufficient to account 
for peculiar velocities at the boundary of our mock catalog.

Peculiar velocities of subhalos are defined as
the average velocity of particles within the innermost 10\% 
of the virial radius.  
The virial overdensity in {\sf Rockstar} is defined by 
$\Delta_{\rm vir}\approx 237\rho_{\rm m}$ at $z=0.534$.  
This definition does {\em not} correspond to the definition of the halo bulk flow velocity that 
uses all particle members of the halo, because the halo core and its outer regions 
have different velocity structure. For a demonstration of this point, see Figure 11 of \citet{Behroozi_2013a} 
and also Appendix B of R14.\footnote{The definition of halo peculiar velocity in R14 is the average
of particles within $\sim 33\%$ of the virial radius 
where $\Delta_{\rm vir}\approx 200\rho_{\rm m}$.}
All subhalos are mapped into redshift space by including the peculiar velocity component 
along the $Z$ direction before performing abundance matching.

%%%%%%%%%%%%%%%%%%%%%%%%%%%%%%%%%%%%%%%%%%%%%%%%%%%%%%%%%%%%%%%%%%%%%%%%%%%%%%%
\subsection{Time evolution and Resolution Tests}
%%%%%%%%%%%%%%%%%%%%%%%%%%%%%%%%%%%%%%%%%%%%%%%%%%%%%%%%%%%%%%%%%%%%%%%%%%%%%%%

In this section, we discuss potential issues in the subhalo
catalog, focusing in particular on the time evolution of subhalo
clustering and completeness issues due to the resolution of the
simulation. Here we only summarize our findings -- Figures and further
details can be found in Appendix.~\ref{app:subhalotest}.

We adopt a single redshift output at $z=0.534$ from the {\sf MDR1}
simulation. We test if a single redshift output is sufficient
to model CMASS over the redshift range $0.43<z<0.7$. There are three
outputs available in the {\sf MDR1} simulation over the redshift range
of interest: $z=0.466, 0.534$ and $0.609$. Using these redshift
outputs, we find a difference in the real-space correlation function
at fixed number density,
$\overline{n} \simeq 1.58\times 10^{-4} (h^{-1}{\rm Mpc})^{-3}$, at the
$1\operatorname{-}2$ \% level at large scales. The largest differences
(at the level of 5\%) are seen at the 1-halo to 2-halo regime at
$r\lesssim 1\,h^{-1}{\rm Mpc}$ (see Appendix.~\ref{app:subhalotest}). 
This level of evolution is below our measurement errors, but these effects
will need to be taken into account in future work, especially when the
$S/N$ of the measurements increases (currently we are using DR10
measurements).

We also perform two tests concerning the impact of the resolution of
{\sf MDR1} on our results. First, we determine if the subhalo catalog
resolves the mass scale required for our abundance matching. 
Based on \cite{White_2011} and R14, we estimate that abundance matching for CMASS will require
subhalos with $V_{\rm peak}\ge 200\,{\rm km\,s^{-1}}$. Our tests demonstrate
that {\sf MDR1} resolves halos down to $V_{\rm peak}\sim 150\,{\rm km\,s^{-1}}$.

Second, we examine the impact of resolution effects on the radial
profiles of subhalos. Our estimates suggest that subhalo radial
profiles become incomplete at $0.1$-$0.7\,h^{-1}{\rm Mpc}$ (and depend on
the ratio between the peak velocity of hosts and subhalos). The
smallest scale in our $w_{p}$ measurement is
$\approx 0.2\,h^{-1}{\rm Mpc}$ and is close to this incompleteness
limit. The impact of resolution on our results is at least partly counteracted by 
the boost to the errors of our measured $w_{p}$ by systematic fiber-collision correction uncertainties on
these scales. We conclude that the resolution of {\sf MDR1} is
sufficient for our purpose, but that recently-completed higher resolution simulations such as
\citet{Skillman:2014kx} or \citet{Ishiyama:2015} would be preferable
and will be adopted in subsequent work.

%%%%%%%%%%%%%%%%%%%%%%%%%%%%%%%%%%%%%%%%%%%%%%%%%%%%%%%%%%%%%%%%%%%%%%%%%%%%%
%%%%%%%%%%%%%%%%%%%%%%%%%%%%%%%%%%%%%%%%%%%%%%%%%%%%%%%%%%%%%%%%%%%%%%%%%%%%%
\section{Methodology}
\label{sec:methodology}
%%%%%%%%%%%%%%%%%%%%%%%%%%%%%%%%%%%%%%%%%%%%%%%%%%%%%%%%%%%%%%%%%%%%%%%%%%%%%
%%%%%%%%%%%%%%%%%%%%%%%%%%%%%%%%%%%%%%%%%%%%%%%%%%%%%%%%%%%%%%%%%%%%%%%%%%%%%

Our goal is to find a model of the CMASS-halo connection which can
simultaneously explain the SMF and the two-point correlation function
and which also accounts for stellar mass completeness of CMASS. This
section explains the details of our methodology. In this
paper we only explore models that reproduce the projected two-point
correlation function of the full CMASS sample over the redshift range
of $0.43<z<0.7$.  In future work we will explore how well our models
match the clustering of sub-samples (e.g., dividing CMASS by color and
redshift).

We begin with a broad overview of our global methodology and the two
classes of models explored in this paper. The details of
our approach are then provided in the later half of this section. The
casual reader may wish to read the overview of the methodology and then
skip directly to the summary provided in Section \ref{subsec:predictwp}.

%%%%%%%%%%%%%%%%%%%%%%%%%%%%%%%%%%%%%%%%%%%%%%%%%%%%%%%%%%%%%%%%%%%%%%%%%%%%%
\subsection{Overview of Methodology and Models}
\label{Sec:SHAM_Models}
%%%%%%%%%%%%%%%%%%%%%%%%%%%%%%%%%%%%%%%%%%%%%%%%%%%%%%%%%%%%%%%%%%%%%%%%%%%%%

Our approach is based on the SHAM framework for connecting galaxies
and dark matter halos (see Section \ref{subsec:SHAM}). Within the
context of SHAM, we will explore two broad classes of models that
relate galaxy color to halo properties. The first model that we
explore is a ``stochastic model'' in which at fixed stellar mass,
galaxy color in high-mass halos is simply a random process that does
not correlate with halo properties. We will refer to this model as the
``AbM" model. After abundance matching our mock catalog, we {\it tag}
CMASS galaxies by randomly down-sampling the full mock galaxy catalog
in such a way that the mock CMASS SMFs reproduce the ones measured in
Section \ref{sec:CMASS_SMFz}.  Unless an additional correlation
between this CMASS flag and halo properties is explicitly introduced,
this procedure makes the implicit assumption that {\em at fixed
  stellar mass, CMASS galaxies are a random sample of the overall
  population}.  However, L15 show that at fixed stellar mass, CMASS is
not a random sample of the overall population in terms of galaxy
color.  Hence, the abundance matched catalog that we obtain after the
down-sampling procedure will only correctly represent the true
relation between galaxy color, stellar mass, and halo properties if
color is a random process at fixed stellar mass.

The second model is an extension to the traditional abundance
matching scheme introduced by H13 called {\it age matching}. This
model is based on the premise that galaxy color correlates with a
secondary halo property at fixed stellar mass (see Section
\ref{subsec:AgM_Xcol}). After first abundance matching our mock
catalog, the age-matched model will be built by re-shuffling CMASS
galaxies according to a secondary halo property. In order to fully
implement the age-matching model, however, we need to characterize the
color distributions of galaxies from the {\sc s82-mgc} as a function
of mass and redshift and also to understand the effects of scatter
introduced in these color distributions from photometric
redshifts. This is a non-trivial task that we defer to Paper II -
opting here instead to simply perform a {\it qualitative}
investigation of the effects of age matching on the two-point
correlation function. For this purpose, we will adopt a simple color
model for the galaxy population that is based on a ``color-rank
distribution'' represented by $X_{\rm col}$ which effectively
characterizes the color ranking of the CMASS versus other galaxies.
This distribution is characterized by one free parameter called
$\mu_{\rm CMASS}$.  As described in \S\ref{subsec:AgM_Xcol}, this
parameter controls the correlation strength between subhalo properties
and the CMASS selection function.

%%%%%%%%%%%%%%%%%%%%%%%%%%%%%%%%%%%%%%%%%%%%%%%%%%%%%%%%%%%%%%%%%%%%%%%%%%%%%
\subsection{Subhalo Abundance Matching}
\label{subsec:SHAM}
%%%%%%%%%%%%%%%%%%%%%%%%%%%%%%%%%%%%%%%%%%%%%%%%%%%%%%%%%%%%%%%%%%%%%%%%%%%%%

The SHAM scheme provides an effective and simple way to model the
galaxy-halo relation and has been successful at modeling both the
galaxy stellar mass function as well as the galaxy two-point
correlation function \citep[see
e.g.,][]{Kravtsov_2004,Vale_2004,Conroy_2006,Moster_2010,Behroozi_2010}.
The basic philosophy of SHAM is that massive (sub)halos host bright
galaxies.  More concretely, the SHAM method begins by rank ordering
galaxies by stellar mass $M_{*}$ (or luminosity). Halos drawn from
$N$-body simulations are rank ordered by peak maximum circular
velocity $V_{\rm peak}$.  Galaxies are then assigned to subhalos in
descending order such that $n_{\rm gal}(>M_{*})=n_{\rm halo}(>V_{\rm
  peak})$.  In practice, there are multiple ambiguities in the SHAM
technique.  First, there is freedom in choosing how to rank order
subhalos.  For example, \citet{Reddick_2013} showed how the predicted
two point correlation function varies when rank ordering is performed
using different halo mass proxies such as halo mass $M_{\rm vir}$,
maximum circular velocity $V_{\rm circ}$, and its peak over entire
merging history, $V_{\rm peak}$.  Motivated by this work, we will
evaluate how our model varies when rank ordering by either $V_{\rm
  peak}$ or $M_{\rm peak}$.  Second, SHAM models must also account for
scatter between galaxy properties and halo properties.  We account for
scatter by adopting the methodology of \citet{Behroozi_2010} and
\citet{Reddick_2013}.

To perform abundance matching, we need to evaluate the total SMF over
the entire mass range covered by the CMASS sample, i.e., down to
$\log_{10}(M_*/M_{\odot})\sim 10.6$. This value is below the
completeness limit of the {\sc s82-mgc}. Our strategy will be to fit
the total SMF from the {\sc s82-mgc} in the range
$\log_{10}(M_*/M_{\odot})>11.5$ using a double Schechter function \citep{Baldry_2008}:

%%%%%%
%\begin{equation}
\begin{eqnarray}
  &&\phi(M_{*};\phi_{1},\alpha_{1},\phi_{2},\alpha_{2},M_{0}) 
  =  \left\{ \phi_{1}10^{(\alpha_{1}+1)(\log M_{*} -\log M_{0})}\right.\nonumber\\
   &&  \;\;\;\; \left.  + \phi_{2}10^{(\alpha_{2}+1)(\log M_{*} -\log M_{0})} \right\}
    (\ln 10)\exp\left[-\frac{M_{*}}{M_{0}}\right], 
    \label{eq:phi_dSchechter}
\end{eqnarray}
%\end{equation}
%%%%%%
where $|\alpha_{2}|>|\alpha_{1}|$ and the second term dominate at the
low-mass end. The amplitude of the SMF below 
  $\log_{10}(M_*/M_{\odot})=11.5$ is unconstrained by the {\sc
  s82-mgc} SMF but will be adjusted by our joint fit to the clustering
of CMASS galaxies. Section \ref{sec:Results} shows that our joint fit
to the {\sc s82-mgc} SMF and to $w_p$ yields a SMF that is consistent
at the low-mass end with results from PRIMUS and COSMOS.

We abundance match subhalos against this SMF, and
convolve it with a uniform log-normal scatter,
%%%%%%
%\begin{equation}
\begin{eqnarray}
  &&\phi_{\rm conv}(M_{*};\phi_{1},\alpha_{1},\phi_{2},\alpha_{2},M_{0},\sigma)\nonumber\\
  &=& \int\,dm\,\frac{\phi(10^{m})}{\sqrt{2\pi}\sigma}
  \exp\left[-\frac{(m-\log M_{*})^{2}}{2\sigma^{2}}\right], 
\label{eq:phi_conv}
\end{eqnarray}
%\end{equation}
%%%%%%
which introduces a scatter in the relation between stellar and halo mass. 
This scatter arises due to a combination of intrinsic scatter 
in the stellar-to-halo-mass relation and errors associated with stellar mass 
measurements \citep{Behroozi_2010,Leauthaud_2011}. 
Hence, for a realistic model, the value of $\sigma$ must be equal to, or greater 
than, the measurement errors in stellar mass measurements 
-- we will return to this question in Section \ref{sec:Results}. 

We fit the {\sc s82-mgc} SMF over 8 data points at $11.5 \le
\log_{10}(M_*/M_{\odot})\le 12.0$.  Our SMF measurements probe the
high mass end of the stellar mass function and hence are insensitive
to some parameters in the double Schechter function.  For this reason,
in our fits, we simply fix the parameters that 
%\sout{govern the low mass end} 
  is not sensitive to the very high mass end to
$(\alpha_{1},\phi_{2},\alpha_{2})=(-0.46, 3.0\times 10^{-4}, -1.58)$.
This is motivated by results at the low-mass end from
\citet{Baldry_2008}.
%Finally, as discussed in \ref{Sec:totalSMF}, we also impose priors on the total SMF 
%based on PRIMUS data in the intermediate mass range 
%($M_{*}= 10^{10.55}M_{\odot}$ and $10^{10.95}M_{\odot}$). 
In summary, our abundance matching model contains three free parameters, 
$\phi_{1}$, $M_{0}$, and $\sigma$. 
%Then a total of three SMF parameters 
%($(\phi_{1}, M_{0}, \sigma$) are estimated by minimizing 

We compute a $\chi^{2}$ for the {\sc s82-mgc} SMF as follows:
%%%%%%
%\begin{equation}
\begin{eqnarray}
%\chi^{2}_{\rm SMF} = 
%\sum_{i}\frac{[\phi_{\rm meas}(M_{*,i})-\phi_{\rm conv}(M_{*,i};\phi_{1},M_{0},\sigma)]^{2}}{\Delta \phi_{\rm meas}(M_{*,i})^{2}}, 
&&\chi^{2}_{\rm SMF} = \sum_{ij}[\phi_{\rm meas}(M_{*,i})-\phi_{\rm conv}(M_{*,i};\phi_{1},M_{0},\sigma)]\nonumber\\
&&\;\;\;\; \times{\bf C}^{-1}_{{\rm SMF},ij}[\phi_{\rm meas}(M_{*,j})-\phi_{\rm conv}(M_{*,j};\phi_{1},M_{0},\sigma)]
%\end{equation}
\end{eqnarray}
%%%%%%
where $\phi_{\rm conv}(M_{*};\phi_{1},M_{0},\sigma)$ is the theoretical SMF 
predicted by Equation (\ref{eq:phi_conv}). 

%We here include only the Poisson noise in the error estimate, 
%$\Delta \phi_{\rm meas}(M_{*,i})$, which has a diagonal component 
%only and is a good approximation at high mass end \cite{Smith_2012}.
 
%%%%%%%%%%%%%%%%%%%%%%%%%%%%%%%%%%%%%%%%%%%%%%%%%%%%%%%%%%%%%%%%%%%%%%%%%%%%%
\subsection{Subhalo Age Matching}
\label{subsec:AgM_Xcol}
%%%%%%%%%%%%%%%%%%%%%%%%%%%%%%%%%%%%%%%%%%%%%%%%%%%%%%%%%%%%%%%%%%%%%%%%%%%%%

SHAM essentially specifies the stellar-to-halo mass relation between
galaxies and halos. It is normally assumed that halo mass is the
primary variable on which the galaxy-halo connection depends. However,
in addition to halo mass, halo clustering also depends on other
parameters such as halo age, a phenomenon known as {\it assembly bias}
\citep[see e.g.,][]{Gao_2005,Wechsler_2006,Jing_2007,Gao_2007,Dalal_2008,Li_2008,Lin:2015fk,Miyatake:2015qy}.

H13 introduced an extension to the traditional abundance
matching scheme called {\it age matching} which can reproduce the
color-dependent clustering of the SDSS main galaxy sample \citep[also
see][]{Masaki_2013}.  This method
matches galaxies and halos using both stellar mass as well as galaxy
color. The basic premise of the approach is that redder galaxies are
assigned to older subhalos at fixed stellar mass.

In the age matching scheme, each halo is assigned a characteristic
redshift ($z_{\rm starve}$) computed from halo merger trees.  This
$z_{\rm starve}$ parameter is defined as the maximum of three distinct
age components:

\begin{itemize}
\item $z_{\rm char}$: the earliest redshift at which the most massive progenitor 
of a given subhalo exceeds $M_{h}>10^{12}\,h^{-1}M_{\odot}$. 
For subhalos less massive than $10^{12}\,h^{-1}M_{\odot}$, $z_{\rm char}=z_{\rm obs}$.  
\item $z_{\rm acc}$: the epoch when a subhalo accretes onto a host halo.
For host halos, $z_{\rm acc}=z_{\rm obs}$.
\item $z_{\rm form}$: the epoch defined by 
$z_{\rm form}=c_{\rm vir}/(4.1a_{\rm acc})-1$, 
motivated by the fact that there is a tight correlation between the concentration 
parameter and the epoch when halo growth transits from a fast to slow 
accretion regime \citep{Wechsler_2006}. Note that $a_{\rm acc}=1/(1+z_{\rm acc})$.
\end{itemize}
We adopt $z_{\rm obs}=0.534$ while in the original work of H13, 
$z_{\rm obs}=0$.

There is a critical difference between this work and H13: 
our relevant mass regime ($\log_{10}(M_{*}/M_{\odot})\gtrsim 11$) is much 
higher than that of H13 ($\log_{10}(M_{*}/M_{\odot})\lesssim 11$).
H13 found that $z_{\rm form}$ is the dominant component of
$z_{\rm starve}$ for the SDSS main sample whereas we find that $z_{\rm char}$ 
is the dominant component for CMASS (see Section \ref{sec:Results}). 
This is in keeping with the results shown in Figure 5 of \citet{Hearin_2014}, which demonstrates 
that $z_{\rm char}$ begins to dominate the contribution to $z_{\rm starve}$ for stellar masses 
$\log_{10}(M_{*}/M_{\odot})\gtrsim 11.5$ at $z\sim 0$. 
In our CMASS sample, these higher-mass galaxies dominate the sample, whereas the Main Galaxy Sample 
is dominated by lower-mass galaxies. 
Because of these key differences, the impact of assembly bias in our models will be 
qualitatively different compared to H13 (see Section \ref{sec:Results}).

In Paper II we will use the actual color distributions of massive
galaxies as a function of redshift to perform age matching. Our goal
in this paper, however, is to perform a first qualitative analysis
of the general effects of age matching above collapse mass, a regime
that has not yet been fully investigated. For this purpose, we
introduce a simple color-rank distribution denoted $X_{\rm col}$. 
This color-rank distribution will be used to assign ``colors'' to
CMASS and to non-CMASS galaxies and to perform the color-based rank
ordering in the age matching scheme. 
Our goal is to construct a model that allows for a
simple ``mixing'' between these two populations.

Operationally, we accomplish this mixing with our age matching model as follows. 
First, at each stellar mass we generate a random distribution of $X_{\rm col}$ values. 
Suppose there are $N_{\rm h}$ subhalos in the stellar mass bin, and that the fraction of 
galaxies of this stellar mass that are CMASS-selected is denoted by $f_{\rm CMASS}.$ 
We then draw $f_{\rm CMASS}\times N_{\rm h}$ values from a 
Gaussian distribution of mean $\mu_{\rm CMASS}$ and unit variance; these draws will be the 
``colors" $X_{\rm col}$ of our mock CMASS galaxies. We next draw $(1-f_{\rm CMASS})\times N_{\rm h}$ values from a 
Gaussian distribution of {\em zero} mean and unit variance; these draws will be the 
``colors" $X_{\rm col}$ of our non-CMASS galaxies. We then rank-order the joint collection of the randomly drawn values of $X_{\rm col}.$ 
Subhalos in the same stellar mass bin are rank-ordered by $z_{\rm starve}$. 
In monotonic fashion, the larger $X_{\rm col}$ draws are assigned to 
the subhalos with larger $z_{\rm starve}$ values, and the CMASS-designation associated with 
$X_{\rm col}$ is also assigned to the subhalo, defining the CMASS selection function in the ``AgM" model. 

The value of $\mu_{\rm CMASS}$ determines 
the strength of the correlation between the CMASS selection function and subhalo $z_{\rm starve}$ at fixed stellar mass. 
If $\mu_{\rm CMASS}$ is large (for instance, $\mu_{\rm CMASS}=10$), then $X_{\rm col}-$values with a CMASS-designation 
are always larger than $X_{\rm col}-$values attached to non-CMASS draws, in which case at fixed stellar mass, 
subhalos with the highest $z_{\rm starve}$ are always selected to be CMASS galaxies. 
On the other hand, if $\mu_{\rm CMASS}=0$, the $X_{\rm col}$ distributions of CMASS and non-CMASS draws
are identical, so in this case matching the 
$X_{\rm col}$ and $z_{\rm starve}$ distributions has no impact on the CMASS designation assigned to the subhalos, 
and the CMASS selection function is uncorrelated with  $z_{\rm starve}$ at fixed stellar mass. 
Finally, for intermediate values of $\mu_{\rm CMASS}$ (for instance, $\mu_{\rm CMASS}=0.6$), 
then CMASS and non-CMASS galaxies have overlapping
$X_{\rm col}$ distributions, but CMASS galaxies are ``redder" on average. 
Figure \ref{fig:dist_X} illustrates these concepts.

In our analysis,
$\mu_{\rm CMASS}$ is left as a free parameter which means that we
determine the degree to which CMASS colors overlap with non CMASS
galaxies directly from the data. 
We do not however 
currently account for any redshift and 
stellar-mass dependence 
of $\mu_{\rm CMASS}$, thus we do not account for any redshift and stellar-mass 
dependence of the CMASS color-cuts.
This is a limitation of our current model, the importance of which will 
become clearer in Section \ref{subsec:AgMex}.

%%%%%%%%%%%%%%%%%%%%%%%%%%%%%%%%%%%%%%%%%%%%%%%%%%%%%%%%%%%%%%%%%%%%%%%%%%%%%
\subsection{Accounting for the Stellar Mass Completeness of CMASS as
  a Function of Redshift}
\label{subsec:mstar_incompleteness}
%%%%%%%%%%%%%%%%%%%%%%%%%%%%%%%%%%%%%%%%%%%%%%%%%%%%%%%%%%%%%%%%%%%%%%%%%%%%%

We assume a single global SMF over the CMASS redshift range. For each
set of parameters $(\phi_{1},M_{0},\sigma)$, we
create a mock catalog via abundance matching. At this point galaxies
in the mock catalog have redshifts and stellar masses. The next step
is to tag CMASS galaxies in the abundance-matched mock catalog as a
function of redshift. We divide our simulation into seven redshift bins
along the $Z$ direction (the bin width is $\Delta z=0.04$). 
The redshift width of $\Delta z=0.04$ is conservative and this choice is 
motivated by the uncertainty of photometric redshift estimation in the {\sc s82-mgc}
catalog. 

The CMASS SMF varies as a function of redshift as a result of the BOSS selection function.  
In Section \ref{sec:CMASS_SMFz}, we used the {\sc s82-mgc} 
catalog to measure the number densities of CMASS galaxies
as a function of mass and redshift, $N^{\rm CMASS}_{\rm sim}(M_{*},z)$.  
Because we assume that the total SMF does not vary
over our redshift baseline, we can compute how many CMASS galaxies are
expected for every redshift slice in the mock catalog simply by
scaling this number by the ratio of volume in the redshift slice in
the mock ($\Delta V_{\rm sim}(z)$) to the {\sc s82-mgc} volume
($\Delta V_{\rm S82}(z)$):
%%%%%
\begin{equation}
 N^{\rm CMASS}_{\rm sim}(M_{*},z) = 
 \frac{\Delta V_{\rm sim}(z)}{\Delta V_{\rm S82}(z)}
  N^{\rm CMASS}_{\rm S82}(M_{*},z). 
\end{equation}
%%%%%
In order to predict the number of mock galaxies as a function of mass
and redshift, we construct bins in stellar mass from 10.6 to 12.3 dex
with $\Delta \log M_{*}=0.05$.  We have checked that our prediction is
stable with $\Delta \log M_{*}=0.1$.  In the mock catalog, we randomly
tag $N_{\rm CMASS}(M_{*},z)$ galaxies with a CMASS flag.  For a small
number of bins, $N^{\rm CMASS}_{\rm sim}(M_{*},z)$ exceeds the number
predicted by the total SMF (simply due to sample variance).  In this
case, we simply set $N^{\rm CMASS}=N^{\rm tot}$.  Following this
procedure, every galaxy in our mock catalog is now assigned a stellar
mass, a redshift, and a flag that indicates mock CMASS galaxies.  By
design, mock CMASS galaxies have stellar mass distributions that match
the ones measured in Section \ref{sec:CMASS_SMFz}.

%%%%%%%%%%%%%%%%%%%%%%%%%%%%%%%%%%%%%%%%%%%%%%%%%%%%%%%%%%%%%%%%%%%%%%%%%%%%%
\subsection{Predicting the CMASS Two-Point Correlation Function}
\label{subsec:predictwp}
%%%%%%%%%%%%%%%%%%%%%%%%%%%%%%%%%%%%%%%%%%%%%%%%%%%%%%%%%%%%%%%%%%%%%%%%%%%%%

We now have a mock catalog that contains galaxies with three
dimensional positions and with a flag that indicates CMASS
galaxies. The next step is to compute the predicted the CMASS
two-point correlation function.  $w_{\rm p,theory}$ is computed from
the mock following the exact same procedure as for the BOSS DR10 data.
To account for the finite volume of the simulation, we compute a
covariance matrix for $w_{\rm p,theory}$ 
(referred to as ${\bf C}_{w_{\rm p},{\rm theory}}$), which is estimated via jack-knife by
dividing the $(X,Y)$-plane into 256 equal regions.  For the small scales
of concern in this paper, jack-knife errors outperform
bootstrap errors (P.~Norberg, private communication, 
Arnalte-Mur \& Norberg et al., in prep).

The fitting for $w_{\rm p}$ is performed with
%%%%%%
\begin{equation}
\chi^{2}_{w_{\rm p}} = \sum_{i,j}\Delta_{w_{\rm p}}(r_{{\rm p},i};\phi_{1},M_{0},\sigma){\bf C}_{w_{\rm p},{\rm total},ij}^{-1}\Delta_{w_{\rm p}}(r_{{\rm p},j};\phi_{1},M_{0},\sigma), 
\end{equation}
%%%%%%
where $\Delta_{w_{\rm p}}(r_{{\rm p},i};\phi_{1},M_{0},\sigma)=w_{\rm p,meas}(r_{{\rm p},i})-w_{\rm p,theory}(r_{{\rm p},i};\phi_{1},M_{0},\sigma)$, 
and the total covariance matrix includes uncertainties in 
both measurement and our theory estimates, i.e., 
${\bf C}_{w_{\rm p},{\rm total}}={\bf C}_{w_{\rm p},{\rm meas}}+{\bf C}_{w_{\rm p},{\rm theory}}$. 

%=======================================================================%
\begin{figure}
\begin{center}
\hspace*{-0.7cm}
\includegraphics[width=1.1\columnwidth]{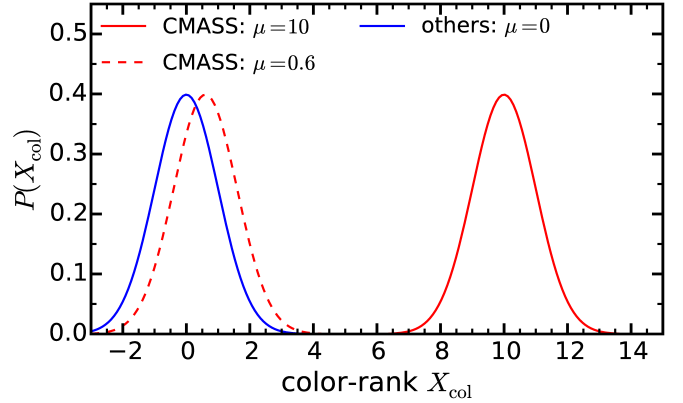}
\caption{\label{fig:dist_X} Illustrative figure of the color-rank 
  distributions for CMASS and non-CMASS galaxies. The $X_{\rm col}$
  ``colors'' of non-CMASS galaxies are drawn from a normal distribution
  with unit variance and zero mean (shown by the solid blue line).
  The $X_{\rm col}$ ``colors'' of CMASS galaxies are drawn from a
  normal distribution with unit variance and with a mean value equal
  to $\mu_{\rm CMASS}$. When $\mu_{\rm CMASS}=0.599$ (dashed red line),
  CMASS and non-CMASS galaxies have overlapping color distributions
  but CMASS is redder on average.  When $\mu_{\rm CMASS}=10$ (solid
  red line), all CMASS galaxies are redder than non-CMASS galaxies
  (this situation corresponds to the extreme age-matching case 
  explored in Section \ref{subsec:AgMex}).  Our best-fitting value for
  $\mu_{\rm CMASS}$ is 0.599 and corresponds to the distribution shown
  by the dashed red line.}
\end{center}
\end{figure}

\begin{figure*}
\begin{center}
\includegraphics[width=1.8\columnwidth]{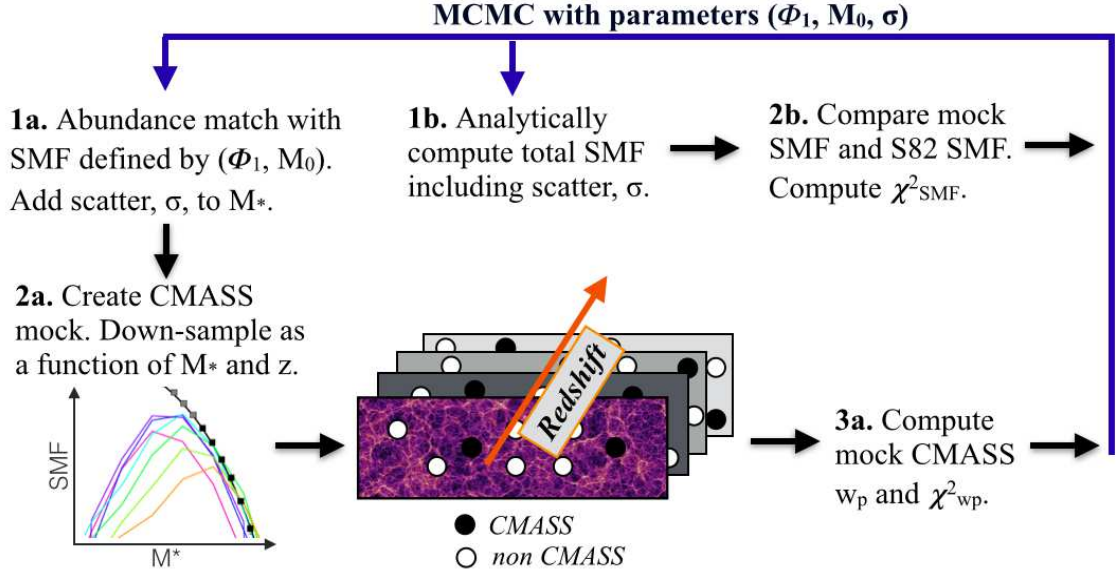}
\caption{\label{fig:methodology} Illustration of our overall
  methodology for constraining the AbM model and creating a mock CMASS
  catalog. The stochastic AbM model contains three free parameters:
  $(\phi_{1},M_{0},\sigma)$. The age-matching (AgM) model contains one
  additional parameter, $\mu_{\rm CMASS}$, which controls how strongly
  CMASS galaxies correlate with $z_{\rm starve}$ 
  at fixed $V_{\rm peak}$.
}
\end{center}
\end{figure*}
%=======================================================================%

%%%%%%%%%%%%%%%%%%%%%%%%%%%%%%%%%%%%%%%%%%%%%%%%%%%%%%%%%%%%%%%%%%%%%%%%%%%%%
\subsection{Summary of Methodology}
\label{subsec:predictwp}
%%%%%%%%%%%%%%%%%%%%%%%%%%%%%%%%%%%%%%%%%%%%%%%%%%%%%%%%%%%%%%%%%%%%%%%%%%%%%

Figure \ref{fig:methodology} presents an illustration of our
methodology for the AbM model. A summary of our methodology is as
follows:
\begin{enumerate}

\item Start with a set of SMF parameter values. For the stochastic
  (``AbM'') model, the parameters are $(\phi_{1},M_{0},\sigma)$.  For
  the age-matching (``AgM'') model there is an 
  additional parameter, $\mu_{\rm CMASS}$.  The parameter $\mu_{\rm
    CMASS}$ only impacts the modeling of two-point statistics such as
  $w_{\rm p}$ - one-point statistics such as the SMF are entirely
  unaffected by $\mu_{\rm CMASS}.$

\item The two parameters $\phi_{1}$ and $M_{0}$ control the total SMF
  (without scatter). The total SMF including scatter is obtained
  analytically following Equation (\ref{eq:phi_conv}). A
  $\chi^{2}_{\rm SMF}$ is computed between this analytic model and the
  total stellar mass function estimated in Section
  \ref{sec:totalSMF}. 

\item In parallel, we generate a mock catalog to model $w_p$.  The
  first step in generating this mock catalog is to abundance match the
  mock catalog using the same total SMF (without scatter) as in the
  previous step.  We test abundance matching both in terms of $V_{\rm
    peak}$ and $M_{\rm peak}$.  Scatter ($\sigma$) is introduced into
  stellar mass in the mock catalog at fixed $V_{\rm peak}$ (or $M_{\rm
    peak}$).  We have checked that the mock catalog is large enough
  that stochasticity due to the rare number of high mass halos is a
  negligible effect, i.e., the Poisson error in the measured mass
  function dominates the error budget at high stellar masses.

\item The $Z$ direction of the mock is taken as the redshift axis.
  Mock CMASS galaxies are tagged in the mock catalog by down-sampling
  the overall population in redshift and stellar mass in order to
  reproduce the CMASS SMFs measured in Section \ref{sec:CMASS_SMFz}.

\item At this stage, mock CMASS galaxies are simply a random
  sub-sample of the overall population; this mock corresponds to
  our stochastic ``AbM" model.
  
\item For the age-matching model, we begin by assigning the subhalos 
  a color-rank, $X_{\rm col}$, as follows. For subhalos hosting a non-CMASS galaxy, 
  $X_{\rm col}$ is drawn from a Gaussian distribution with zero mean and unit variance.
  For subhalos hosting a CMASS galaxy, $X_{\rm col}$ is drawn from a Gaussian
  distribution with mean $\mu_{\rm CMASS}$ and unit variance. At fixed
  stellar mass, the random $X_{\rm col}$ values are rank-ordered. 
  At the same stellar mass, the mock galaxies are rank-ordered by a secondary halo property, 
  where we choose $z_{\rm starve}$ as this
  secondary parameter in our fiducial model, 
  where $z_{\rm starve}$ is concretely defined in \S \ref{subsec:AgM_Xcol}. 
  In monotonic fashion, the subhalos with the largest $z_{\rm starve}$ 
  values are assigned the largest $X_{\rm col}$ values, and the CMASS/non-CMASS designation 
  associated with each $X_{\rm col}$ value is also assigned to the subhalo.\footnote{For certain tests, 
  we may rank-order the subhalos only according to $z_{\rm form}$ or $z_{\rm char}$ (Section
  \ref{subsec:AgM_Xcol}).} Thus subhalos in the ``AbM" and ``AgM" mocks in general have different 
  CMASS-designations: in ``AgM", the CMASS-designation is correlated
  with $z_{\rm starve}$ at fixed $V_{\rm peak}$, with the correlation strength 
  governed by our $\mu_{\rm CMASS}$ parameter. 

%\item To generate the age-matched mock, galaxies are assigned a random
%  color rank, $X_{\rm col}$. For non-CMASS galaxies, $X_{\rm col}$ is
%  drawn from a Gaussian distribution with zero mean and unit variance.
%  For mock CMASS galaxies, $X_{\rm col}$ is drawn from a Gaussian
%  distribution with mean $\mu_{\rm CMASS}$ and unit variance. At fixed
%  stellar mass (or $V_{\rm peak}$), galaxies are rank-ordered by
%  $X_{\rm col}$ and reshuffled by a secondary halo property. For our
%  fiducial age matching mock we adopt $z_{\rm starve}$ as this
%  secondary parameter.  For certain tests, we may reshuffle only
%  according to $z_{\rm form}$ or $z_{\rm char}$ (Section
%  \ref{subsec:AgM_Xcol}).

\item Generate a random catalog that follows the CMASS $dn/dz$ and
  compute $w_{\rm p,theory}$. Note that ${\bf C}_{\rm theory}$ is
  fixed using our best-fitting parameters (after a first initial
  iteration).

\item Compute $\chi^2_{\rm w_{\rm p}}$ between $w_{\rm p,meas}$ and
  $w_{\rm p,theory}$, and then add as $\chi^{2}=\chi^2_{\rm
    SMF}+\chi^2_{\rm w_{\rm p}}$.

\item Iterate this procedure. 
\end{enumerate}

The best-fit parameters and errors are determined using the Markov
chain Monte Carlo (MCMC) technique.  We use a modified version of
COSMOMC \citep{Lewis_2002} that has been well tested in previous work
\citep{Saito_2011,Zhao_2013,Saito_2014}.  Since our SMF is
estimated from {\sc s82-mgc} and the correlation function $w_{\rm p}$ is
computed over the full DR10 footprint, the cross correlation between these
two statistics are negligible.  The $\chi^{2}$ for the multipoles is
defined in a similar way. The correlation between the monopole and the
quadrupole is properly taken into account by the covariance matrix.

%In previous work, \citep[][]{Nuza_2013} also
%modeled the CMASS two-point correlation function using
%SHAM. Improvements that we have implemented with respect to
%\citep[][]{Nuza_2013} include the fact that a) our abundance matching is
%performed with respect to 
%and b) we account for the stellar mass 
%completeness of the CMASS data as a function of redshift. 

%%%%%%%%%%%%%%%%%%%%%%%%%%%%%%%%%%%%%%%%%%%%%%%%%%%%%%%%%%%%%%%%%%%%%%%%%%%%%%%%%%
%%%%%%%%%%%%%%%%%%%%%%%%%%%%%%%%%%%%%%%%%%%%%%%%%%%%%%%%%%%%%%%%%%%%%%%%%%%%%%%%%%
\section{Results}
\label{sec:Results}
%%%%%%%%%%%%%%%%%%%%%%%%%%%%%%%%%%%%%%%%%%%%%%%%%%%%%%%%%%%%%%%%%%%%%%%%%%%%%%%%%%
%%%%%%%%%%%%%%%%%%%%%%%%%%%%%%%%%%%%%%%%%%%%%%%%%%%%%%%%%%%%%%%%%%%%%%%%%%%%%%%%%%

%%%%%%%%%%%%%%%%%%%%%%%%%%%%%%%%%%%%%%%%%%%%%%%%%%%%%%%%%%%%%%%%%%%%%%%%%%%%%%%%%%
\subsection{Abundance Matching}
\label{subsec:AbM}
%%%%%%%%%%%%%%%%%%%%%%%%%%%%%%%%%%%%%%%%%%%%%%%%%%%%%%%%%%%%%%%%%%%%%%%%%%%%%%%%%%

\begin{figure*}
\begin{center}
\includegraphics[width=2\columnwidth]{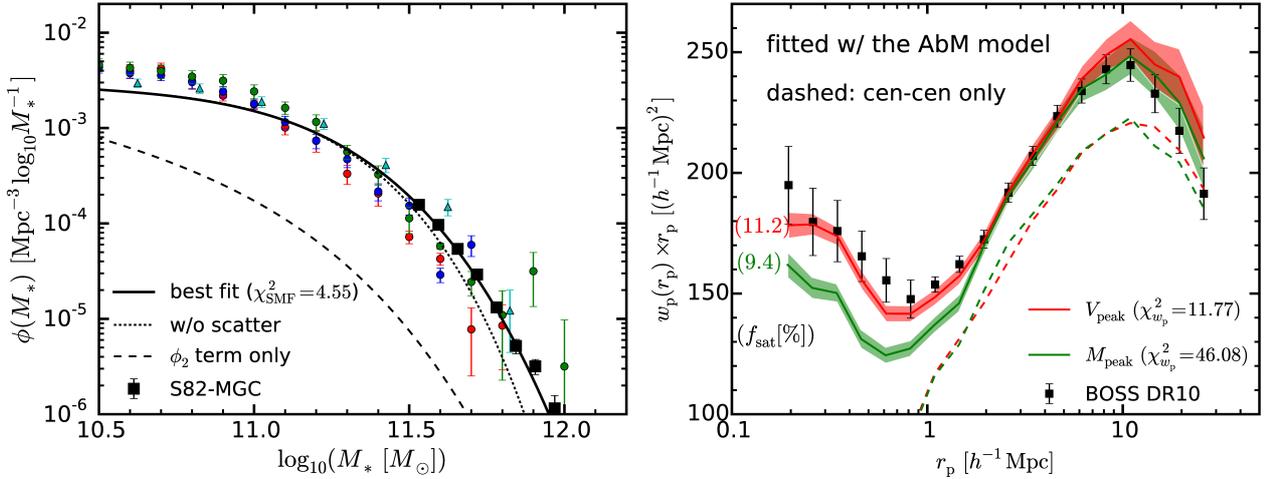}
\caption{\label{fig:SMF_wp_jointfit} {\it Left}): best fit to the
  {\sc s82-mgc} SMF for the AbM model (solid black line). The dotted
  black line corresponds to the SMF deconvolved for scatter. The black
  dashed curved shows the (fixed) $\phi_{2}$ term in our double
  Schechter function. Black squares correspond to the measured SMF from
  the {\sc s82-mgc}. ({\it Right}): our best fit to $w_{\rm p}$ for
  the AbM model (solid red line). The green line shows the result of
  abundance matching against $M_{\rm peak}$ instead of $V_{\rm peak}$.
  Dashed lines display the contribution to $w_{\rm p}$ from
  central-central pairs. Numbers in parenthesis indicate satellite
  fractions (11.1\% for $V_{\rm peak}$ and 9.5\% for $M_{\rm peak}$).
  The goodness of fit for the AbM model is
  $\Delta \chi^{2}=(4.55+11.77)/(26-3)=0.710$.
  }
\end{center}
\end{figure*}

We now perform a joint fit to the SMF and to $w_{\rm p}$. The left
panel of Figure \ref{fig:SMF_wp_jointfit} presents our best-fit to the
SMF using a double Schechter function and abundance-matching against
$V_{\rm peak}$. The best-fit parameters for the double Schechter
function are:
$(\phi_{1},\log_{10}M_{0},\sigma)=(1.86^{+0.21}_{-0.61}\times
10^{-3},10.89^{+0.05}_{-0.04},0.105^{+0.024}_{-0.032})$
with $\chi^{2}_{\rm SMF}=4.55$. Errors are reported with a 68\%
confidence level. We find excellent fits to both the SMF and
$w_{\rm p}$ with two specific points worth highlighting.  First, the
amplitude of our best fit SMF agrees well with COSMOS and PRIMUS at
$\log_{10}M_{*}\gtrsim 11.0$ but has a lower amplitude at
$\log_{10}M_{*}\lesssim 11.0$.  Because the number density of CMASS
drops sharply below this mass scale, we simply do not expect to
constrain the total SMF in this range. Second, the best fit value for
the scatter is lower than our naive expectation.  Indeed, $\sigma$
should include contributions from measurements errors as well as from
intrinsic scatter in the stellar-to-halo mass relation. The average
uncertainty in stellar mass measurements from the {\sc s82-mgc} is of
order $\sigma_{\rm meas}\sim0.1$ dex in this mass and redshift
range. Hence, a value of $\sigma=0.105$ implies a very small 
{\it intrinsic} scatter in the stellar-to-halo mass relation.  We will
return to this point in the discussion section.

The right hand panel of Figure \ref{fig:SMF_wp_jointfit} presents our
best-fit to $w_{\rm p}$ as the red line ($\chi^{2}_{w_{\rm p}}=11.43$). 
The goodness of fit in this case is $\chi^{2}/({\rm d.o.f.})=(4.55+11.77)/(8+18-3)=0.710$.
We have also tested abundance matching against $M_{\rm peak}$ instead
of $V_{\rm peak}$. The blue line shows the results of abundance
matching against $M_{\rm peak}$ using the same best-fitting SMF
parameters as above.  As can be seen from Figure
\ref{fig:SMF_wp_jointfit}, $V_{\rm peak}$ yields a larger clustering
amplitude and is more consistent with the BOSS data than
$M_{\rm peak}$.  

There are two factors which lead to the differences in these clustering predictions. 
First, host halos selected by $M_{\rm peak}$ cluster more weakly relative to host halos 
selected by $V_{\rm peak},$ as discussed in \citet{Zentner_2014}. This effect is small in our 
halo mass range, which we have verified by comparing the central-central pair counts between 
the two models, which are nearly identical. 
Second, the satellite fractions predicted by the two models are different: our $V_{\rm peak}-$based 
SHAM model has a larger satellite fraction relative to the $M_{\rm peak}-$based model. Indeed, at fixed
$M_{\rm peak}$, subhalos have larger $V_{\rm peak}$ than host halos
\citep[see Figure 2 in][]{Reddick_2013}, which suggests that
rank-ordering with $V_{\rm peak}$ results in similar clustering of
central galaxies but the larger satellite fraction boosts the overall
clustering amplitude. We adopt $V_{\rm peak}$ as our fiducial model
and do not explore abundance matching with $M_{\rm peak}$ any further.

%%%%%%%%%%%%%%%%%%%%%%%%%%%%%%%%%%%%%%%%%%%%%%%%%%%%%%%%%%%%%%%%%%%%%%%%%%%%%%%%%%
\subsection{CMASS $dn/dz$}
\label{subsec:cmassdndz}
%%%%%%%%%%%%%%%%%%%%%%%%%%%%%%%%%%%%%%%%%%%%%%%%%%%%%%%%%%%%%%%%%%%%%%%%%%%%%%%%%%

Figure \ref{fig:CMASS_dndz} presents a comparison between the redshift
distribution of CMASS galaxies from our best-fitting mock catalog with
the redshift distribution of CMASS galaxies in the {\sc s82-mgc} and
from the full BOSS DR10 SGC. Our mock reproduces the CMASS $dn/dz$
from the {\sc s82-mgc} catalog and is consistent with $dn/dz$ from the
BOSS DR10 SGC. The amplitude differences between the $dn/dz$ from our
mock and the DR10 $dn/dz$ are due to sample variance. In our current
methodology, the sample variance introduced by matching the CMASS SMFs
from Stripe 82 is not taken into account which is a limitation of our
current approach. This reflects a trade-off made to take advantage of
the higher quality stellar mass estimates from the {\sc s82-mgc}, but
doing so, our current analysis is also limited by the sample variance
from Stripe 82.

%%%%%%%%%%%%%%%%%%%%%%%%%%%%%%%%%%%%%%%%%%%%%%%%%%%%%%%%%%%%%%%%%%%%%%%%%%%%%%%%%%
\subsection{Gaining an Intuition for Age Matching Above Collapse Mass}
\label{subsec:AgMex}
%%%%%%%%%%%%%%%%%%%%%%%%%%%%%%%%%%%%%%%%%%%%%%%%%%%%%%%%%%%%%%%%%%%%%%%%%%%%%%%%%%

In the previous section, we showed that a reasonable fit to 
$\Phi$ and $w_{\rm p}$ can be achieved using a simple abundance matching scheme 
in which galaxy color in high mass halos is simply a stochastic process. 
We now investigate whether or not models in which galaxy color 
{\em correlates} with halo assembly properties can achieve comparable results. 
One caution worth mentioning here with respect to the age matching model is that, unlike in H13, 
the combination of the steep $V_{\rm peak}{\mathchar`-}M_{*}$ relation
and the non-zero scatter in this relation leads to a difference in the mean host halo 
mass compared to the standard abundance matching model (see Appendix. 
\ref{app:scatter_and_agm} for details). 

First, we wish to develop some intuition for how the
different components of $z_{\rm starve}$ affect $w_{\rm p}$ in this
high halo mass regime.  Figure \ref{fig:histogram_Mvir} shows that
CMASS galaxies are firmly in halo masses above collapse mass,
$M_{\rm col}(z=0.534)=10^{11.73}\,h^{-1}{\rm M_{\odot}}$.  Hence, the
behavior of $z_{\rm starve}$ may be fundamentally different compared
to previous work by H13. Let us begin by considering an extreme case
in which CMASS galaxies are all redder than non-CMASS galaxies
($X_{\rm col,CMASS}\gg X_{\rm col,others}$, see solid lines in Figure
\ref{fig:dist_X}). For this test, we adopt the values of the best fit
to $\Phi$ from Figure \ref{fig:SMF_wp_jointfit} . At fixed stellar
mass, we rank order galaxies according to $z_{\rm starve}$,
$z_{\rm form}$, or $z_{\rm char}$. The results of this extreme case
are presented in Figure \ref{fig:AgM_wp}.  Interestingly, but perhaps
not surprisingly, we find that $z_{\rm form}$ (blue curve) 
{\em lowers} the clustering amplitude.  This is because $z_{\rm form}$ is
defined using halo concentrations and the effects of assembly bias
have an opposite effect above and below collapse mass when using the
concentration parameter \citep[see
e.g.][]{Wechsler_2006,Dalal_2008}. Thus, in this high-mass regime,
$z_{\rm form}$ causes red galaxies to cluster {\em less} strongly than
blue ones.

Let us now turn our attention to $z_{\rm char}$. Interestingly, rank
ordering according to $z_{\rm char}$ produces the opposite effect and
causes an {\em increase} in the clustering amplitude.  Previous work
on assembly bias has shown the switch in the assembly bias effect seen
when considering halo concentration is not always reflected when
considering other halo parameters. Previous work has not studied the
specific case of $z_{\rm starve}$; however, \citet{Jing_2007} and
\citet{Li_2008} report that when an age parameter based on a fixed mass
threshold such as $z_{1/2}$ is used 
where $z_{1/2}$ denotes the redshift when a halo acquires half of the final 
mass at the observational time,  
a similar behavior is observed
\citep[see Figure 4 in][]{Li_2008}.

Finally, let us now examine the $z_{\rm starve}$
component, which includes contributions from both
$z_{\rm char}$ and $z_{\rm form}$.  The prediction for
$z_{\rm starve}$ lies between the $z_{\rm form}$ and the
$z_{\rm char}$ cases but is closer to $z_{\rm char}$ than to
$z_{\rm form}$. This is because in this mass regime, $z_{\rm starve}$
is dominated by $z_{\rm char}$ not $z_{\rm form}$ (see
Figure \ref{fig:zstarve_contribution}). Thus the impact of the
assembly bias for CMASS is qualitatively distinct from the
trends identified by H13 in lower mass halos, a fact which traces to the change in character 
of assembly bias for halos above and below collapse mass. 

The dashed curves in Figure \ref{fig:AgM_wp} display $w_{\rm p}$ for
central galaxies only -- demonstrating that the trends discussed above
%are indeed two-halo assembly bias effects and 
are not simply due to varying satellite fractions. 

%%%%%%%%%%%%%%%%%%%%%%%%%%%%%%%%%%%%%%%%%%%%%%%%%%%%%%%%%%%%%%%%%%%%%%%%%%%%%%%%%%
\subsection{Fit to $w_p$ with an Age Matching Type Model}
\label{subsec:AgMfit}
%%%%%%%%%%%%%%%%%%%%%%%%%%%%%%%%%%%%%%%%%%%%%%%%%%%%%%%%%%%%%%%%%%%%%%%%%%%%%%%%%%

Of course, the true differences between the color distributions of
CMASS galaxies compared to non-CMASS galaxies of similar mass are not
as extreme as the case explored in the previous section. As discussed
in section \ref{Sec:SHAM_Models}, the implementation of age-matching
first requires a characterization of the color distributions of
galaxies from the {\sc s82-mgc} as a function of mass and redshift and
also requires modeling the effects of scatter introduced in these
color distributions from photometric redshifts. This is an aspect that
we defer to Paper II. Here, we perform a {\it qualitative}
investigation of the effects of age matching on the two-point
correlation function using the color-rank variable
$\mu_{\rm CMASS}$.

We now perform a joint fit to the SMF and to $w_{\rm p}$ in which
$\mu_{\rm CMASS}$ is left as a free parameter (the ``AgM'' model). The
results are presented in Figure \ref{fig:FittedAgM_wp}. The best-fit
parameters are
$(\phi_{1},\log_{10}M_{0},\sigma,\mu_{\rm
  CMASS})=(2.51^{+1.71}_{-0.75}\times
10^{-3},10.83^{+0.04}_{-0.11},0.136^{+0.38}_{-0.22},
0.599^{+0.435}_{-0.164})$
with $\chi^{2}_{\rm SMF}=4.09$ and $\chi^{2}_{w_{\rm p}}=10.75$. The
AbM model and the AgM model yield a comparable goodness of fit
($\Delta \chi^{2}=(4.55+11.77)/(26-3)=0.710$ for the AbM model and
$\Delta \chi^{2}=(4.09+10.75)/(26-4)=0.674$ for the AgM model). There
are three points worth highlighting concerning the results of the AgM
model. First, the best-fit SMF has a slightly higher amplitude at low
stellar masses compared to the AbM model and is in better agreement
with PRIMUS and COSMOS. Second, the best-fit value for the scatter
is larger than the AbM model ($\sigma=0.136$ versus
$\sigma=0.105$), which leaves a larger margin for intrinsic
scatter. Finally, the best-fit value for $\mu_{\rm CMASS}$ of
$0.599$ corresponds to a scenario in which CMASS and non-CMASS
galaxies have overlapping color distributions, but with CMASS galaxies
being somewhat redder on average. Reassuringly, this result matches our
qualitative expectations for this sample.

The AgM model explored here is simplistic in the sense that we have
used a single value of $\mu_{\rm CMASS}$ over a whole CMASS redshift
range, whereas the true color distribution of CMASS versus other
galaxies depends on redshift.  A more sophisticated model which
accounts for this effect will be presented in a forthcoming paper.

\begin{figure}
\begin{center}
\hspace*{-0.7cm}
\includegraphics[width=1.1\columnwidth]{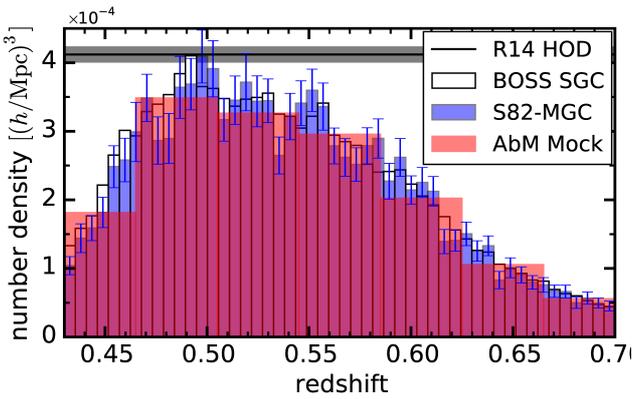}
\caption{\label{fig:CMASS_dndz} Comparison between the CMASS $dn/dz$
  from our fiducial mock catalog (red histograms), the measured
  $dn/dz$ from the {\sc s82-mgc} (blue histograms), and the measured
  $dn/dz$ from the full BOSS DR10 SGC (white histograms, \citet{Anderson_2014}). 
  Errors on the $dn/dz$ for the {\sc s82-mgc} are
  estimated via bootstrap. For the DR10 SGC $dn/dz$, redshift failures
  and fiber-collided galaxies are included using a nearest-neighbor
  weighting scheme (see \citet{Anderson_2014}). By construction, our
  models reproduce the redshift distribution of CMASS galaxies from
  the {\sc s82-mgc} catalog which is in turn consistent with the DR10
  SGC CMASS redshift distribution. The number density from the
  fiducial R14 model is shown as a horizontal solid black line. In the
  R14 model, the CMASS $dn/dz$ is reproduced by randomly down-sampling
  a fixed redshift independent HOD.}
\end{center}
\end{figure}

\begin{figure}
\begin{center}
\hspace*{-1.2cm}
\includegraphics[width=1.1\columnwidth]{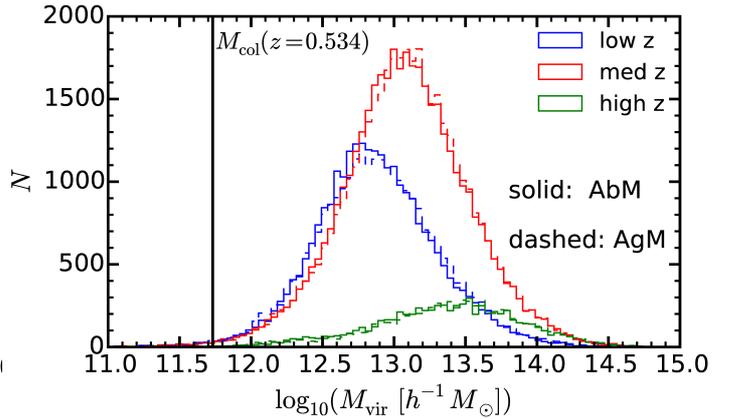}
\caption{\label{fig:histogram_Mvir} Halo mass histograms as a function
  of redshift from our AbM (solid lines) and AgM (dashed lines) mock
  catalogs. Collapse mass at $z=0.534$ is indicated by a black solid
  vertical line.  Clearly, CMASS galaxies populate halos with masses
  firmly above collapse mass. Also note that the mean halo mass of
  CMASS in our mocks varies by a factor of 3.5 from low to high redshift.}
\end{center}
\end{figure}

\begin{figure}
\begin{center}
\hspace*{-1cm}
\includegraphics[width=1.1\columnwidth]{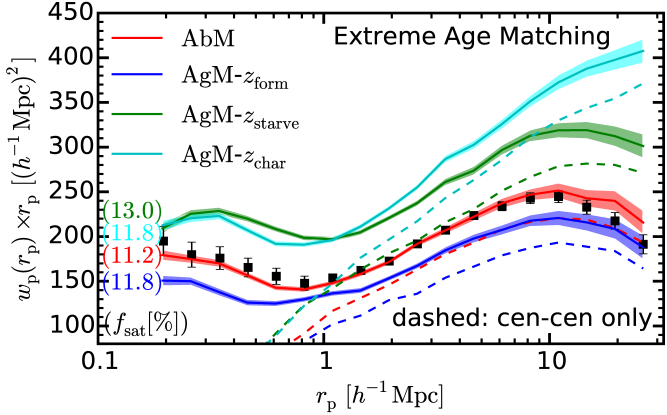}
\caption{\label{fig:AgM_wp} Impact of age matching (AgM model) on
  $w_{\rm p}$ for an extreme scenario with $\mu_{\rm CMASS}=10$ (CMASS
  galaxies are redder than all other galaxies). Rank ordering is
  performed versus $z_{\rm form}$ (blue), $z_{\rm starve}$ (green) and
  $z_{\rm char}$ (cyan). For comparison, we also present the best-fit
  curves from the AbM model (red solid line) in which the correlation
  between the colors of CMASS galaxies and subhalo age is completely
  stochastic. This goal of this figure is simply to highlight the
  qualitative trends of age-matching above collapse mass. Rank
  ordering versus $z_{\rm char}$ increases the amplitude of
  $w_{\rm p}$ whereas rank ordering versus $z_{\rm form}$ decreases
  the amplitude of $w_{\rm p}$. Dashed lines show the contribution to
  $w_{\rm p}$ from central-central pairs.  Numbers in parenthesis
  indicate satellite fractions. %
}
\end{center}
\end{figure}

\begin{figure}
\begin{center}
\hspace*{-0.7cm}
\includegraphics[width=1.1\columnwidth]{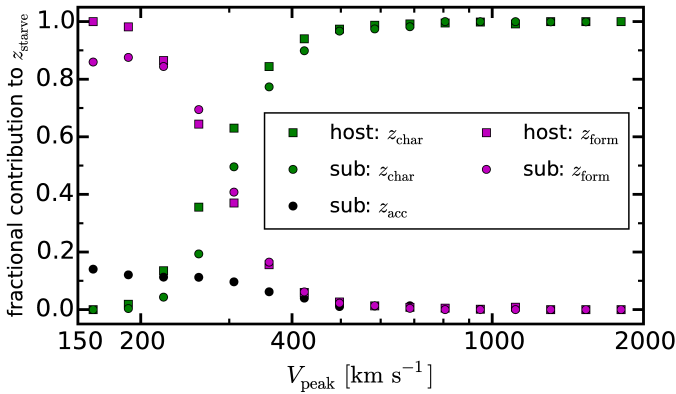}
\caption{\label{fig:zstarve_contribution} Fractional contribution to
  $z_{\rm starve}$ as a function of $V_{\rm peak}$ at $z=0.534$ for
  host ({\it square}) and sub ({\it circle}) halos. The $z_{\rm char}$
  term dominates at the high mass end whereas the $z_{\rm form}$ term
  dominates at the low mass end.%
}
\end{center}
\end{figure}

%\begin{figure}
%\begin{center}
%\includegraphics[width=1\columnwidth,bb=0 0 522 328]{fig/wp_fittedAgM.pdf}
%\caption{\label{fig:FittedAgM_wp} 
%Results when age matching (AgM) is performed with $z_{\rm starve}$ and
%$\mu_{\rm CMASS}$ is adjusted to fit to the BOSS DR10 data (blue line). 
%We find a best-fit value of $\mu_{\rm CMASS}=0.5$. The red lines shows the  ABM result.
%Dashed lines show the contribution to wp from central-central pairs. 
%Numbers in parenthesis indicate satellite fractions.
%
%}
%\end{center}
%\end{figure}

\begin{figure*}
\begin{center}
\includegraphics[width=2\columnwidth]{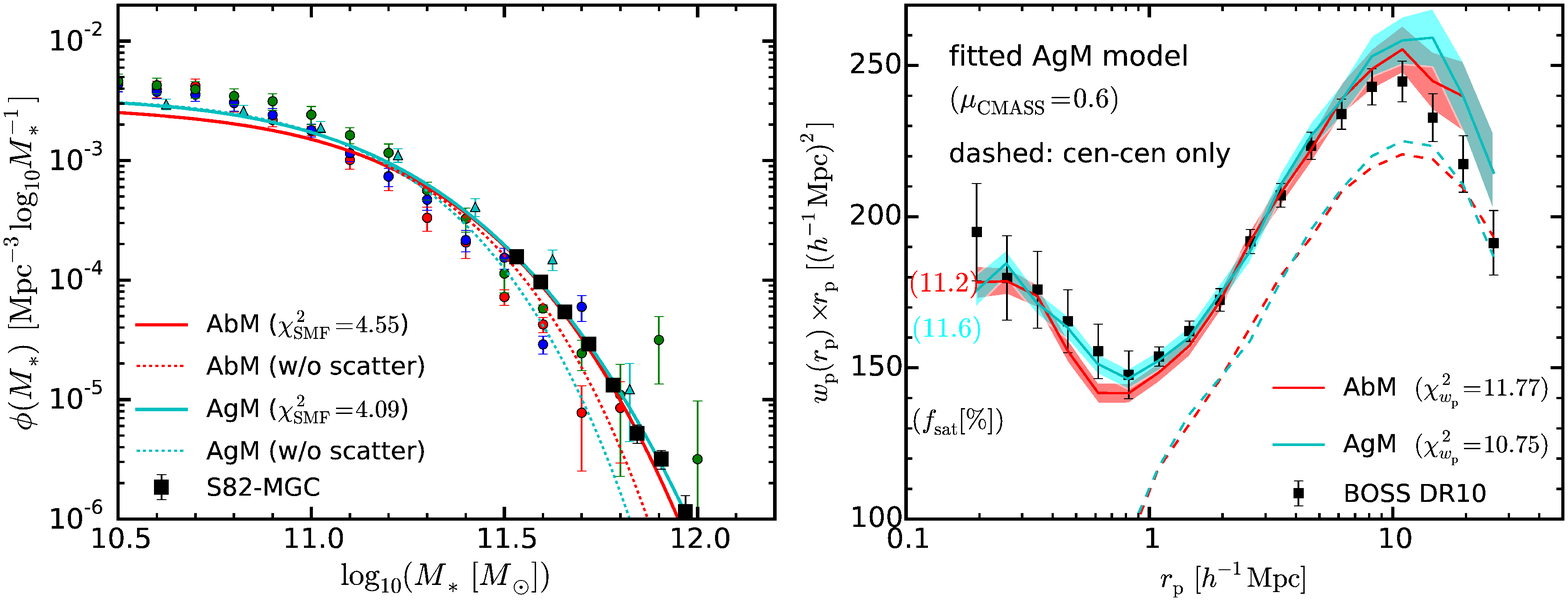}
\caption{\label{fig:FittedAgM_wp} {\it Left}): our best fit to the {\sc s82-mgc}
  SMF for the AgM model (solid cyan line). The dotted cyan line
  corresponds to the SMF deconvolved for scatter. For comparison, the
  AbM result is displayed with red lines. Black squares correspond to
  the measured SMF from the {\sc s82-mgc}. ({\it Right}): our best fit
  to $w_{\rm p}$ for the AgM model (solid cyan line). For comparison,
  the AbM result is shown as a red solid line. Dashed lines display
  the contribution to $w_{\rm p}$ from central-central pairs. Numbers
  in parenthesis indicate satellite fractions. The goodness of fit for
  the AgM model is $\Delta \chi^{2}=(4.09+10.75)/(26-4)=0.674$. }
\end{center}
\end{figure*}

%=======================================================================%

%%%%%%%%%%%%%%%%%%%%%%%%%%%%%%%%%%%%%%%%%%%%%%%%%%%%%%%%%%%%%%%%%%%%%%%%%%%%%%%%%%
%%%%%%%%%%%%%%%%%%%%%%%%%%%%%%%%%%%%%%%%%%%%%%%%%%%%%%%%%%%%%%%%%%%%%%%%%%%%%%%%%%
\section{Discussion}
\label{sec:Discussion}
%%%%%%%%%%%%%%%%%%%%%%%%%%%%%%%%%%%%%%%%%%%%%%%%%%%%%%%%%%%%%%%%%%%%%%%%%%%%%%%%%%
%%%%%%%%%%%%%%%%%%%%%%%%%%%%%%%%%%%%%%%%%%%%%%%%%%%%%%%%%%%%%%%%%%%%%%%%%%%%%%%%%%

%%%%%%%%%%%%%%%%%%%%%%%%%%%%%%%%%%%%%%%%%%%%%%%%%%%%%%%%%%%%%%%%%%%%%%%%%%%%%%%%%%
\subsection{HOD Modeling in the Context of Complex Samples such as CMASS}
\label{subsec:HOD}
%%%%%%%%%%%%%%%%%%%%%%%%%%%%%%%%%%%%%%%%%%%%%%%%%%%%%%%%%%%%%%%%%%%%%%%%%%%%%%%%%%

Both HOD and SHAM are popular methods for modeling the SMF and the
galaxy-two-point correlation functions. One reason that HOD methods
are popular is that they provide a relatively simple framework that
can also be used to rapidly model a variety of observables. However,
one of the downsides of this method is that specific functional forms
must be assumed for the central and satellite occupation functions.
These assumptions may be robust for volume-limited threshold samples
such as those commonly studied in the SDSS main samples 
\citep[e.g.,][]{Zehavi_2011}.  However, it is less clear if these types of
methods can be applied to samples such as CMASS which are
selected via complex color and luminosity cuts and for which both the
shape and normalization of the effective HOD may vary with redshift.

There have been several attempts to model the CMASS-halo connection on
the basis of HOD type models. Among these studies, 
\citet{Guo_2013,Guo_2014} and \citet{More:2014qy} focused on
specific sub-samples of CMASS, whereas \citet{White_2011} and R14 used
a HOD type model to describe the clustering of the full CMASS,
assuming no redshift evolution in the HOD.

In this paper, we have introduced a novel SHAM-based method\footnote{After this paper was submitted, 
a parallel effort was brought to our attention 
which adopts a similar methodology as our paper \citep[][]{Rodriguez-Torres:2015kx}.}
that can be used to model complex populations such as CMASS by accounting for 
the mass completeness of the sample as a function of redshift. 
We explore a first qualitative approach for also considering color completeness 
which will be developed further in Paper II. 
We now investigate what these models predict in terms of the
redshift dependence of the CMASS HOD. The right panel of Figure
\ref{fig:TOTAL_SMF} shows that the SMF of CMASS varies strongly with
redshift. This figure alone suggests that the HOD of the CMASS sample
is not likely to be uniform over the CMASS redshift range. 

\begin{figure*}
\begin{center}
\includegraphics[width=2.1\columnwidth]{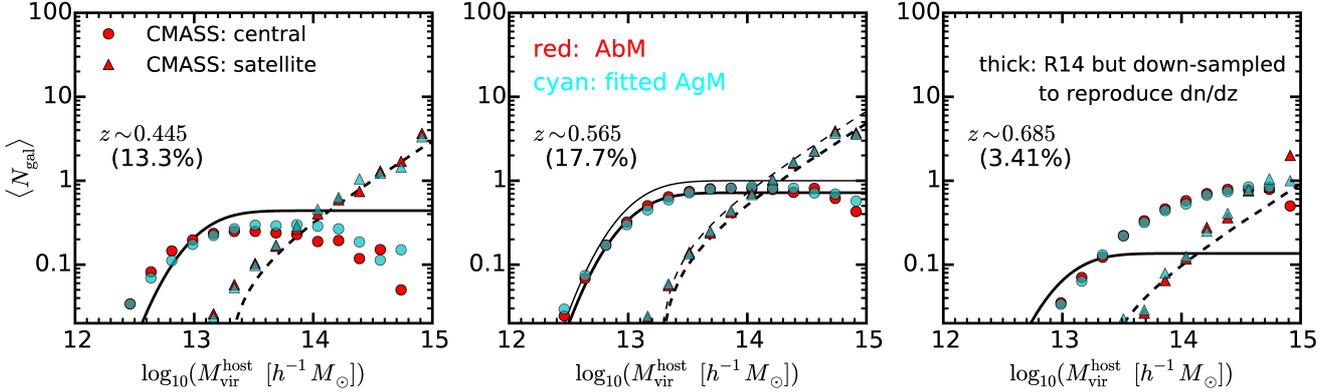}
\caption{\label{fig:HOD} Redshift dependent CMASS HODs from our AbM
  (red circles and triangles) and AgM (cyan circles and triangles)
  mock catalogs. The thin black lines in the middle panel correspond
  to the fiducial R14 CMASS HOD. 
  Note that the virial halo mass in R14 is converted to the Rockstar one. 
  The solid line represents centrals
  and the dashed line represents satellites. Our models should be
  compared with the thick black lines which correspond to the R14
  CMASS HOD after down-sampling to match the CMASS $dn/dz$. Numbers in
  parenthesis represent the percentage of CMASS galaxies in each
  redshift bin compared to the full sample. 
  The data of the HOD table as a function of redshift will be 
  made publicly available at \url{www.massivegalaxies.com}.
  }
\end{center}
\end{figure*}

Figure \ref{fig:HOD} presents the HODs predicted from our AbM and AgM
mock catalogs as a function of redshift. As a comparison we also
display the HOD from R14, which assumes no redshift evolution. 
R14 fit the clustering assuming a constant number density 
with a derived value of $\overline{n}=(4.12\pm 0.13)\times 10^{-4}\,(h^{-1}{\rm Mpc})^{-3}$ 
(see Figure \ref{fig:CMASS_dndz}) under the assumption that the CMASS $dn/dz$ can
be obtained by simply down-sampling the best-fit HOD as a function of
redshift. We down-sample the R14 HOD to match the CMASS $dn/dz$ and
present the results in Figure \ref{fig:HOD}.

There are several noteworthy differences between our HODs and the
single non-evolving one from R14.  At the lowest redshift bin,
$z=0.445$, the mean occupation for central galaxies does not approach
unity due to incompleteness in the SMF at high mass end (see the
magenta curve in the right panel of Figure \ref{fig:TOTAL_SMF}).  At
$z=0.565$, which corresponds to the peak of the CMASS $dn/dz$, our HOD
is more similar to R14, but there is still a discrepancy in the shape
of $\langle N_{\rm cen}\rangle$, especially at the low mass end. The
largest differences are at $z>0.6$.  Our HODs converge to unity at
large halo masses whereas the down-sampled R14 one converges to
$N_{\rm tot}\sim0.1$; this is due to the stellar mass completeness of
CMASS. This difference arises, because, in our models, the
decline of the CMASS number density above $z=0.55$ is caused by the
fact that the mean stellar mass of the sample increases (as
constrained by data from the {\sc s82-mgc}). In contrast, the fixed
HOD of R14 must significantly down-sample the overall amplitude of the
HOD to achieve comparable number densities.

\begin{figure}
\begin{center}
\includegraphics[width=1\columnwidth]{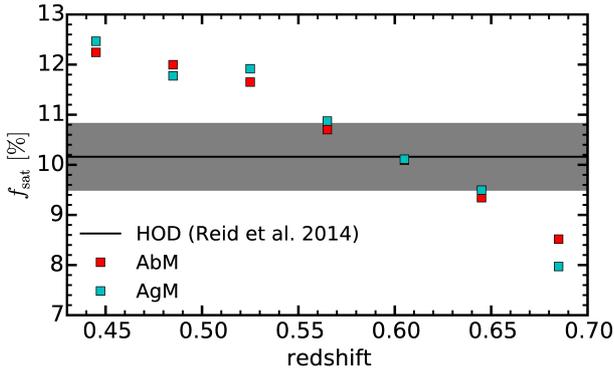}
\caption{\label{fig:CMASS_fsat} Redshift evolution of the satellite
  fraction predicted from our AbM (red squares) and AgM (blue squares)
  models. The redshift independant satellite fraction from R14 is
  shown as a horizontal black line. The grey shaded region indicates
  the $1\sigma$ error on the R14 satellite fraction. The satellite
  fraction in our SHAM models evolves with redshift and is only
  consistent with R14 at $z\sim 0.6$.}
\end{center}
\end{figure}

Finally, our model predicts an evolution of the mean halo mass of
CMASS, as a function of redshift. More specifically, our models predict
that, at $z=0.445,0.565$ and $0.685$, the mean halo mass of central
CMASS galaxies is $\log_{10}(\overline{M}_{\rm halo}\,[h^{-1}M_{\odot}])=$ 
13.12 (13.15), 13.34 (13.35), and 13.66 (13.68) for 
the abundance-matched (age-matched) cases, respectively.  
This variation is driven by the fact that mean
stellar mass of the sample varies with redshift, as is clearly seen in the
right panel of Figure \ref{fig:TOTAL_SMF}. These values are compared
with the HOD result, $\log_{10}(\overline{M}_{\rm halo}\,[h^{-1}M_{\odot}])=13.51$, 
which is higher (lower) than our results at low (high) redshift.  
In addition, our models predict that the CMASS satellite fraction varies with 
redshift from 12\% to 9\%, as shown in Figure \ref{fig:CMASS_fsat}.  
While this effect might seem like a small and negligible variation, the fiducial HOD 
from R14 constrains the satellite fraction at 6.8 \% precision. It is interesting
that the value inferred from the single HOD fit in R14 is
consistent with our values at $z\sim0.6$ but that not at lower
redshifts.

In conclusion, our work suggests that CMASS is a complex sample for
which the HODs are likely to vary with redshift in a non-trivial
manner. A single HOD fit to the overall $w_{\rm p}$ broadly agrees
with the predictions from our model at the median redshift of the
sample. However, at lower and higher redshifts, our model predicts
that HODs are not simple down-sampled versions of the HOD at the peak
of the $dn/dz$.

%%%%%%%%%%%%%%%%%%%%%%%%%%%%%%%%%%%%%%%%%%%%%%%%%%%%%%%%%%%%%%%%%%%%%%%%%%%%%%%%%%
\subsection{A Cautionary Tale of Modeling Small Scale Statistics}
%%%%%%%%%%%%%%%%%%%%%%%%%%%%%%%%%%%%%%%%%%%%%%%%%%%%%%%%%%%%%%%%%%%%%%%%%%%%%%%%%%

Many previous studies have used a combination of galaxy abundances and the
projected galaxy two-point correlation function in order to constrain
the galaxy-halo connection \citep[e.g.,][]{Leauthaud_2011,Coupon_2015,Zu2015}.
However, just because SHAM or HOD models can reproduce these
observables does not necessarily imply that the models accurately
capture the true underlying galaxy halo connection, i.e., 
just because the model provides a good fit to the data does not
necessarily imply that the model is correct. A clear illustration of
this statement in the context of mock galaxy samples with strong assembly bias
is discussed in \citet{Zentner_2014}. In this paper, we have studied
two distinct models: standard abundance matching and a simplified form
of age matching, abbreviated by AbM and AgM, respectively. We have
demonstrated that both models can reproduce the galaxy SMF as well as
$w_{\rm p}$, suggesting that there are fundamental degeneracies among
traditional HOD model, AbM, and AgM models, in modeling the
SMF and $w_{\rm p}$.  This naturally leads to two interesting and
inter-related questions.

\begin{enumerate}
\item How well do these models predict other statistics derived from the data?
\item Are there other statistics which can distinguish between these two distinct models?
\end{enumerate}

Instead of considering just the projected correlation function, 
we turn our attention to the multipoles of the full 2D correlation
function. Figure \ref{fig:xihatell} shows the pseudo multipoles 
(see Section \ref{sec:meas_stat}) for our best-fitting AbM and AgM models.
The left panel of Figure \ref{fig:xihatell} demonstrates that 
{\em both models fail dramatically to reproduce the pseudo multipoles}
even though both models provide a satisfactory description of $w_{\rm p}$. 
In the following section, we will use the redshift dependent
clustering of CMASS to argue that in addition to stellar mass, galaxy
color must play an important role in determining the clustering of
CMASS galaxies and that the failure of our model in reproducing the
pseudo-multipoles must be a consequence of these effects.

In conclusion, our paper provides a clear cautionary example of the
limitation of inferring the galaxy-halo connection from the projected
correlation function alone.  It is also clear from Figure
\ref{fig:xihatell} that the pseudo-multipoles contain additional
information not captured by $w_{\rm p}$ and that these may represent a
powerful and under-utilized tool to provide additional constraints on
the galaxy-halo connection. These aspects will be explored in greater
detail in a forthcoming paper.

%%%%%%%%%%%%%%%%%%%%%%%%%%%%%%%%%%%%%%%%%%%%%%%%%%%%%%%%%%%%%%%%%%%%%%%%%%%%%%%%%%
\subsection{Redshift Evolution of CMASS Clustering}
%%%%%%%%%%%%%%%%%%%%%%%%%%%%%%%%%%%%%%%%%%%%%%%%%%%%%%%%%%%%%%%%%%%%%%%%%%%%%%%%%%

As discussed in section \ref{subsec:HOD}, one major difference between
the R14 model and this work is the treatment of the redshift
evolution of the CMASS sample. In R14, CMASS is assumed to be a single
homogenous sample with a $dn/dz$ that is modeled by down-sampling a
redshift independent HOD. In contrast, in this paper, the varying
number density of CMASS is a direct result of the measured mass
incompleteness of the sample as constrained by the {\sc 82-mgc}
catalog. We now explore the consequences of these differences by
examining the redshift dependent clustering of CMASS.

The original motivation for the non-evolving HOD in R14 originates
from the observation that the clustering of CMASS galaxies does not
vary strongly with redshift. This is shown by Figure A1 in R14
(reproduced here in the right two panels of Figure \ref{fig:xihatell}). 
Because randomly downsampling galaxies does not modify their clustering, 
the R14 model leads to a constant clustering amplitude with redshift,
which indeed, seems well supported by Figure \ref{fig:xihatell}. 
However, another consequence of this procedure is that the halo mass of 
the CMASS sample is constant with redshift in the R14 model. In contrast, the
{\sc S82-MCG} catalog shows that the {\em stellar mass} of the CMASS
sample increases by a factor of 1.8 over the range $0.43<z<0.7$ which
leads to a factor of 3.5 increase in the predicted mean halo mass of
CMASS based on our SHAM modeling.

How much redshift evolution should we expect in the clustering of
CMASS galaxies given this factor of 1.8 increase in stellar mass? The
right hand side of Figure \ref{fig:xihatell} presents the predicted redshift evolution
of the pseudo-multipoles from our SHAM modeling. We find that the
observed stellar mass variation of the CMASS sample should lead to
more than a factor of 1.5 increase in the clustering amplitude over the
CMASS redshift range\footnote{
%Since we only use a single redshift 
%snapshot of the MDR1 simulation, this factor can be overestimated by 
%the evolution of halo distribution. However, a factor of 1.5 is a result when  
%the difference in $\sigma_8$ is factored out. 
Notice that we here ignore the redshift evolution in the {\sf MDR1} simulation.  
However, as seen in Figure. \ref{fig:Xi_hh_evolution}, the effect of the redshift 
evolution is at the level of 5\%. 
}. 
Figure \ref{fig:xihatell} clearly 
reveals a fundamental contrast between the measured non evolution of the
clustering of CMASS and the expectation based on the redshift-dependent 
stellar mass distributions. This discrepancy is
qualitatively insensitive to the exact details of our SHAM
methodology. The observed increase in stellar mass will lead to a
roughy similar increase in halo mass (and hence clustering amplitude)
independently of the exact halo parameter ($V_{\rm peak}$ or $M_{\rm peak}$)
used in the abundance matching.

We argue that the discrepancy revealed in Figure \ref{fig:xihatell} suggests, 
in addition to stellar mass, {\it galaxy color must also play an important
role in determining the clustering amplitude of CMASS galaxies}. 
Figure 5 in L15 demonstrates that CMASS galaxies display a
range in star-formation histories at fixed stellar mass. At low
redshift and fixed stellar mass, the CMASS selection function excludes
galaxies that have experienced recent star formation. At higher
redshifts ($z>0.6$), the CMASS sample is mainly flux limited and
includes a larger range of galaxy colors at fixed magnitude. A variety
of lensing and clustering studies suggest that, for low mass galaxies,
the clustering of blue galaxies is lower than red galaxies at fixed
stellar mass \citep[e.g.,][]{Tinker_2013}. It is not trivial that these trends persist in
this very high galaxy mass regime, but if so, the inclusion of bluer
galaxies in CMASS at higher redshifts may exactly compensate for the
increase in the mean stellar mass. In other words, {\it the observed
constant clustering of CMASS may be due to a coincidental compensation
between color and stellar mass with redshift.}

%%%%%%%%%%%%%%%%%%%%%%%%%%%%%%%%%%%%%%%%%%%%%%%%%%%%%%%%%%%%%%%%%%%%%%%%%%%%%%%%%%
\subsection{What Determines Color in the Most Massive Galaxies?}
%%%%%%%%%%%%%%%%%%%%%%%%%%%%%%%%%%%%%%%%%%%%%%%%%%%%%%%%%%%%%%%%%%%%%%%%%%%%%%%%%%

One of the main goals of this paper is to understand the connection
between halo properties and the colors of very massive galaxies.  As
shown in Figure \ref{fig:histogram_Mvir}, CMASS galaxies live in halos
with halo masses above $10^{12}\,M_{\odot}$.  In this regime, gas
accretion is thought to be dominated by the ``hot halo mode'' and
heated by pressure-supported shocks to a temperature that limits
star-formation \citep{Dekel_2006}.  In addition, at these halo masses,
``maintenance mode'' feed-back mechanisms, such as radio-mode feed-back, 
are thought to further limit star-formation in the most massive
galaxies \cite{Croton_2006}.  However, massive galaxies at these
redshifts are observationally not all red and dead. The CMASS sample
in fact contains a blue population \citep{Guo_2013,Ross_2013}. Based
on high-resolution Hubble Space Telescope imaging,
\citet{Masters_2011} estimate that $\sim 25\%$ of the CMASS sample has
a late-type morphology associated with the star-forming disk
\citep{Masters_2011}. Using a maximum likelihood approach that
accounts for photometric errors as well as the CMASS selection cuts,
\citet{Montero-Dorta-2014} estimate that 37\% of CMASS object may
intrinsically belong to the blue cloud.

Semi-analytic models (SAMs) sometimes assume that galaxy color in high
mass halos is a stochastic process. For example,
\citet{Lu_2014} adopts a simplified halo quenching model to mimic the
effects of AGN feedback that stops radiative cooling in high-mass
halos. In this model, radiative cooling is randomly switched off when
halos reach a critical mass of $10^{12}\,M_{\odot}$ (with a Gaussian
spread of $\sim$ 0.3 dex). In \citet{Benson_2012}, the 
{\sc galacticus} model is more sophisticated and follows the growth and
spins of black holes. The AGN jet power is computed from the accretion
rates and spins of the black hole and is used to counterbalance
radiative cooling in the hot halo. The parameters of the {\sc galacticus} 
model are tuned to produce a transition around few
$10^{12}\,M_{\odot}$ in halo mass, such that quenching begins above
that mass. In this sense, quenching will be stochastic at $M_{\rm halo}\gtrsim 10^{12}\,M_{\odot}$
but also depends on the black hole accretion rate and spin. In the
{\sc galacticus}  model, feedback may also shut down temporarily, for example
after a merging event with high accretion rates which causes the black
hole accretion disk to transition to a thin (radiative) mode with
weaker jet power.

It is thus interesting to ask what drives color in massive galaxies
which live in halos above $10^{12}\,M_{\odot}$. Is color a
stochastic processes that is simply linked to small episodic amounts
of gas cooling and/or merging events? Or is the color in massive
galaxies linked to halo properties such as halo age and hence perhaps
more fundamentally tied to the large scale reservoir of fuel and the
assembly history?

Our current paper does not fully account for the color selection of
CMASS but we can address some of the questions above. In our
AbM model, galaxy color is randomly assigned at fixed stellar mass. In
the AgM model, on the other hand, color is correlated with
$z_{\rm starve}$ and hence with subhalo age. The degree to which color
correlates with $z_{\rm starve}$ is left as a free parameter and
determined from the data. We have shown that both models can reproduce
the galaxy SMF as well as $w_{\rm p}$ but fail to match the
pseudo-multipoles.

\begin{figure*}
\begin{center}
\hspace*{-0.7cm}
\includegraphics[width=2.2\columnwidth]{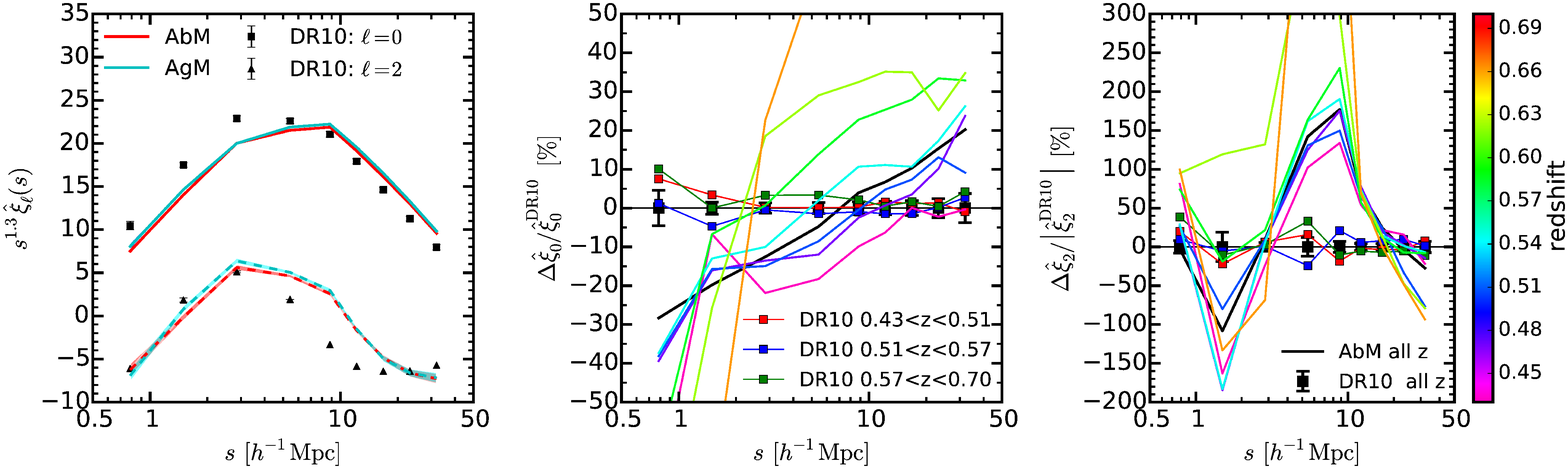}
\caption{\label{fig:xihatell} Left panel: Comparison between the
  measured CMASS pseudo multipoles from R14 and the prediction from
  our AbM and AgM mock catalogs. Solid lines correspond to the pseudo
  monopole and dashed lines correspond to the pseudo
  quadrupole. Neither the AbM or the AgM model are able to reproduce
  the BOSS measurements. 
   Note that our errors on the pseudo multipoles looks smaller than ones in 
  R14 because only measurement errors are included here. 
  Middle panel: redshift evolution of 
  pseudo multipoles 
  in the AbM model prediction are 
  shown as a fractional difference with respect to
  the measurement for the full sample. Red, blue, and green squares
  correspond to BOSS measurements in three different redshift
  bins. The measured BOSS pseudo multipoles display almost no
  variation with redshift.  In stark contrast with the BOSS
  measurements, our models (solid colored lines) predict a significant
  evolution in the pseudo multipoles, driven by the fact that the mean
  stellar mass of CMASS increases by a factor of 1.8 over the range
  $0.43<z<0.7$.
%Middle panel: The fractional difference of the pseudo monopole from the BOSS DR10 data. 
%Our abundance-matching prediction is shown as a red curve, while the best-fitting 
%result from HOD in R14 is shown as a black one. 
%To understand the variations from different velocity effects, we simultaneously 
%plot the lines from HOD by varying three different velocity parameters:
%the mean infall velocity of central galaxies ($\propto f\sigma_{8}$, blue lines), 
%the velocity dispersion of satellite galaxies ($\gamma_{\rm IHV}$, green lines), 
%and the additional dispersion of central galaxies ($\gamma_{\rm cenv}$, magenta line).
%Right panel: Same as the middle panel, but the pseudo quadrupole. 
}
\end{center}
\end{figure*}

To begin, let us focus on the consequences of Figure \ref{fig:xihatell} in
terms of the AbM model. Because the mean stellar mass of the CMASS
sample increases with redshift, the AbM model predicts a strong
variation in the clustering amplitude with redshift which is clearly
ruled out by the data. Hence, we argue that {\it the stochastic color model
(i.e., the AbM model) can be ruled out with high significance by our
analysis.}

We now turn our attention to the AgM model.  Our current
implementation of the AgM model provides an excellent description of
the SMF and $w_p$ but fails to reproduce the
pseudo-multipoles. However, unlike in the case of the AbM model in
which the redshift dependence of the color cuts are unimportant, we
{\em know} that our AgM model will be sensitive to these effects which
we have treated in a simplistic fashion. In a forthcoming paper, we
will investigate if a more realistic AgM model which accounts for the
color completeness of CMASS with redshift can describe the redshift
dependent multipoles. This approach should provide with powerful
constraints on the physical mechanism that drives galaxy color in
massive halos.

%=======================================================================%

%=======================================================================%

%%%%%%%%%%%%%%%%%%%%%%%%%%%%%%%%%%%%%%%%%%%%%%%%%%%%%%%%%%%%%%%%%%%%%%%%%%%%%%%%%%
%%%%%%%%%%%%%%%%%%%%%%%%%%%%%%%%%%%%%%%%%%%%%%%%%%%%%%%%%%%%%%%%%%%%%%%%%%%%%%%%%%
\section{Summary and Conclusions}
\label{sec:summary}
%%%%%%%%%%%%%%%%%%%%%%%%%%%%%%%%%%%%%%%%%%%%%%%%%%%%%%%%%%%%%%%%%%%%%%%%%%%%%%%%%%
%%%%%%%%%%%%%%%%%%%%%%%%%%%%%%%%%%%%%%%%%%%%%%%%%%%%%%%%%%%%%%%%%%%%%%%%%%%%%%%%%%

The last decade has seen rapid observational progress in our
understanding of the relationship between galaxies and their
underlying dark matter halos, However, the connection between galaxies
and dark matter remains poorly constrained for massive galaxies with
$\log_{10}M_{*}\gtrsim 11.5$ because these galaxies are rare with low
number densities, and require large areas surveys to obtain
statistically significant samples.  The BOSS survey provides a
spectroscopic data set of massive galaxies at intermediate redshifts
with number densities of $\bar{n}\approx 3\times 10^{-4}\,[(h^{-1}{\rm Mpc})^{-3}]$ 
in a survey volume that covers several cubic Gigaparsec
(the ``CMASS'' sample). This gigantic dataset enables
high-signal-to-noise ratio measurements of three dimensional galaxy
clustering of massive galaxies.

In this paper, we introduce a novel method based on the SHAM framework
that can be used to model complex populations such as CMASS by
accounting for stellar mass (and eventually color) completeness as a
function of redshift. CMASS is referred to as a ``constant stellar
mass'' sample but L15 demonstrate that CMASS is only
truly stellar mass limited in a narrow mass and redshift range.  In
order to fully utilize this sample to understand the galaxy-halo
connection, it is critical to account for the CMASS mass completeness
function. Our paper accounts for these effects and hence addresses an
important limitation of the CMASS sample which has typically been
neglected in previous work. Our mock catalogs account for CMASS
selection effects, reproduce the overall SMF, the two-point
correlation function of CMASS, and the CMASS $dn/dz$; 
the HOD table as a function of redshift; 
all made publicly available at \url{www.massivegalaxies.com}.
After submitting our paper, a related effort by \citet{Rodriguez-Torres:2015kx}  
was brought to our attention. Several key differences between \citet{Rodriguez-Torres:2015kx}
and our work include the choice of the input stellar mass function, as well as the methodology 
for introducing scatter between stellar and halo mass. 

We use data from Stripe 82 to measure the total SMF down to
$\log_{10}M_{*}\gtrsim 11.5$ and perform a joint fit to both the SMF
and the projected two point correlation function of CMASS
galaxies. Our SHAM model (our ``AbM model") provides an
excellent description of these two observables. Previous work has
assumed that the CMASS HOD does not evolve with redshift.  We
re-investigate this assumption and show that the CMASS HOD should in
fact vary strongly with redshift. Our model predicts that both the
mean halo mass and the CMASS satellite fraction should vary with
redshift. This variation is driven by the fact that the mean stellar mass
of the sample increases at higher redshifts. In conclusion, our work
suggests that CMASS is a complex sample for which the HODs are likely
to vary with redshift in a non-trivial manner.

The color selection applied to the CMASS sample may cause the
two-point correlation function to be sensitive to assembly bias
effects. We study the impact of such effects on the two-point
correlation function using the age matching framework recently
introduced by H13. In contrast with H13, our sample lies firmly above
collapse mass at $z\sim0.55$, which corresponds to a relatively
unexplored mass range. We demonstrate that in this regime, the effects
of assembly bias are markedly different compared to the ones explored
by H13 at lower stellar masses. For example, unlike H13, in this
regime $z_{\rm starve}$ is dominated by $z_{\rm char}$ and not by
$z_{\rm form}$. Also, the $z_{\rm form}$ component of $z_{\rm starve}$
causes red galaxies to cluster {\em less} strongly than blue
ones. However, we also find that the rank ordering according to
$z_{\rm starve}$ produces the opposite effect and causes an {\em
  increase} in the clustering amplitude. We show that an excellent fit
to the CMASS two-point correlation function (which includes assembly
bias effects) can be achieved by balancing these two opposing effects.

Overall, our two distinct models (standard abundance
matching and age matching) can reproduce the galaxy SMF as well as
$w_{\rm p}$, suggesting at first view a fundamental degeneracy between
these models. However, we show that {\em both models fail to reproduce
  the pseudo multipoles} even though both models provide a
satisfactory description of $w_{\rm p}$. Hence, our paper provides a
clear cautionary example of the limitation of inferring the
galaxy-halo connection from the projected correlation function
alone. 

We investigate the redshift dependent clustering of CMASS and find
that the observed stellar mass variation of the CMASS sample should
lead to more than a factor of 2.0 increase in the clustering amplitude
over the CMASS redshift range which is in stark contrast with the
data. We argue that this discrepancy suggests that, in addition to
stellar mass, galaxy color must also play an important role in
determining the clustering amplitude of CMASS galaxies and that the
observed constant clustering of CMASS may be due to a coincidental
compensation between color and stellar mass with redshift. 
Given a discrepancy in shape of the multipole correlation function, it
may be necessary to consider velocity bias as recently 
studied in \citet{Reid_2014,Guo_2014} in the HOD framework. 
However, the velocity bias between subhalos and galaxies are not well 
investigated yet for the mass scale and redshift range of our interest,   
and we defer this aspect to future work (but see \citet{Guo:2016aa} for 
such an effort against the SDSS main sample). 

Finally, we discuss the physical processes that drive galaxy color in
high mass halos. We are interested in determining if color in these
massive galaxies is a stochastic processes that is simply linked to small
episodic amounts of gas cooling and/or merging events. Or is color in
massive galaxies linked to halo properties such as halo age and hence
perhaps more fundamentally tied to the large scale reservoir of fuel
and the assembly history? The stochastic scenario corresponds to our
AbM model in which galaxy color is randomly assigned at fixed stellar
mass. Because the comparison of the redshift dependent clustering of
CMASS with our AbM model, we argue that the stochastic color model can
be ruled out with high significance by our analysis. In this case,
color in high mass halos may be linked to other properties besides
halo peak velocity, suggesting that assembly bias effects may play
a role in determining the clustering properties of this sample.

Our current implementation of age-matching also fails to reproduce the
pseudo-multipoles. However, unlike in the case of the AbM model in
which redshift dependence of the color cuts are unimportant, 
we {\em know} that our AgM model will be sensitive to these effects which we
have treated in a simplistic fashion. Hence, in a forthcoming paper,
we will characterize the CMASS color distributions in greater detail
and investigate if a more realistic age-matching model can describe
the CMASS pseudo-multipoles. This approach will provide powerful
constraints on the physical mechanisms that drives galaxy color in
massive halos.

%%%%%%%%%%%%%%%%%%%%%%%%%%%%%%%%%%%%%%%%%%%%%%%%%%%%%%%%%%%%%%%%%%%%%%%%%%
% ACKNOWLEDGMENTS
%%%%%%%%%%%%%%%%%%%%%%%%%%%%%%%%%%%%%%%%%%%%%%%%%%%%%%%%%%%%%%%%%%%%%%%%%%
\vspace{0.5cm}

We are grateful to Francisco Prada, Risa Wechsler, Chiaki Hikage, and
Surhud More for useful discussions.  We acknowledge Yu Lu and Andrew
Benson for useful discussions related to SAMs.  This work was
supported by World Premier International Research Center Initiative
(WPI Initiative), MEXT, Japan.  Numerical computations were partly
carried out on Cray XC30 at Center for Computational Astrophysics,
National Astronomical Observatory of Japan.  S.S. is supported by a
Grant-in-Aid for Young Scientists (Start-up) from the Japan Society
for the Promotion of Science (JSPS) (No. 25887012). 
The MultiDark Database used in this paper and the web application 
providing online access to it were constructed as part of the activities of 
the German Astrophysical Virtual Observatory as result of a collaboration 
between the Leibniz-Institute for Astrophysics Potsdam (AIP) and the Spanish MultiDark 
Consolider Project CSD2009-00064. The Bolshoi and MultiDark simulations were 
run on the NASA's Pleiades supercomputer at the NASA Ames Research Center. 
The MultiDark-Planck (MDPL) and the BigMD simulation suite have been performed 
in the Supermuc supercomputer at LRZ using time granted by PRACE.
Funding for SDSS-III has been provided by the Alfred P. Sloan Foundation, 
the Participating Institutions, the National Science Foundation, and the
U.S. Department of Energy Office of Science. The SDSS-III web site is
\url{http://www.sdss3.org/}. SDSS-III is managed by the Astrophysical
Research Consortium for the Participating Institutions of the SDSS-III
Collaboration including the University of Arizona, the Brazilian
Participation Group, Brookhaven National Laboratory, Carnegie Mellon
University, University of Florida, the French Participation Group, the
German Participation Group, Harvard University, the Instituto de
Astrofisica de Canarias, the Michigan State/Notre Dame/JINA
Participation Group, Johns Hopkins University, Lawrence Berkeley
National Laboratory, Max Planck Institute for Astrophysics, Max Planck
Institute for Extraterrestrial Physics, New Mexico State University,
New York University, Ohio State University, Pennsylvania State
University, University of Portsmouth, Princeton University, the
Spanish Participation Group, University of Tokyo, University of Utah,
Vanderbilt University, University of Virginia, University of
Washington, and Yale University.

%%%%%%%%%%%%%%%%%%%%%%%%%%%%%%%%%%%%%%%%%%%%%%%%%%%%%%%%%%%%%%%%%%%%%%%%%%
% APPENDIX
%%%%%%%%%%%%%%%%%%%%%%%%%%%%%%%%%%%%%%%%%%%%%%%%%%%%%%%%%%%%%%%%%%%%%%%%%%

\appendix
%%%%%%%%%%%%%%%%%%%%%%%%%%%%%%%%%%%%%%%%%%%%%%%%%%%%%%%%%%%%%%%%%%%%%%%%%%%%%%%%%%
%%%%%%%%%%%%%%%%%%%%%%%%%%%%%%%%%%%%%%%%%%%%%%%%%%%%%%%%%%%%%%%%%%%%%%%%%%%%%%%%%%
\section{Tests of the subhalo catalog}
\label{app:subhalotest}
%%%%%%%%%%%%%%%%%%%%%%%%%%%%%%%%%%%%%%%%%%%%%%%%%%%%%%%%%%%%%%%%%%%%%%%%%%%%%%%%%%
%%%%%%%%%%%%%%%%%%%%%%%%%%%%%%%%%%%%%%%%%%%%%%%%%%%%%%%%%%%%%%%%%%%%%%%%%%%%%%%%%%

In this appendix, we discuss potential issues in the subhalo catalog, 
focusing in particular on the time evolution of subhalo clustering and 
completeness issues due to the resolution of the simulation.\par 

\begin{figure}
\begin{center}
\includegraphics[width=1\columnwidth]{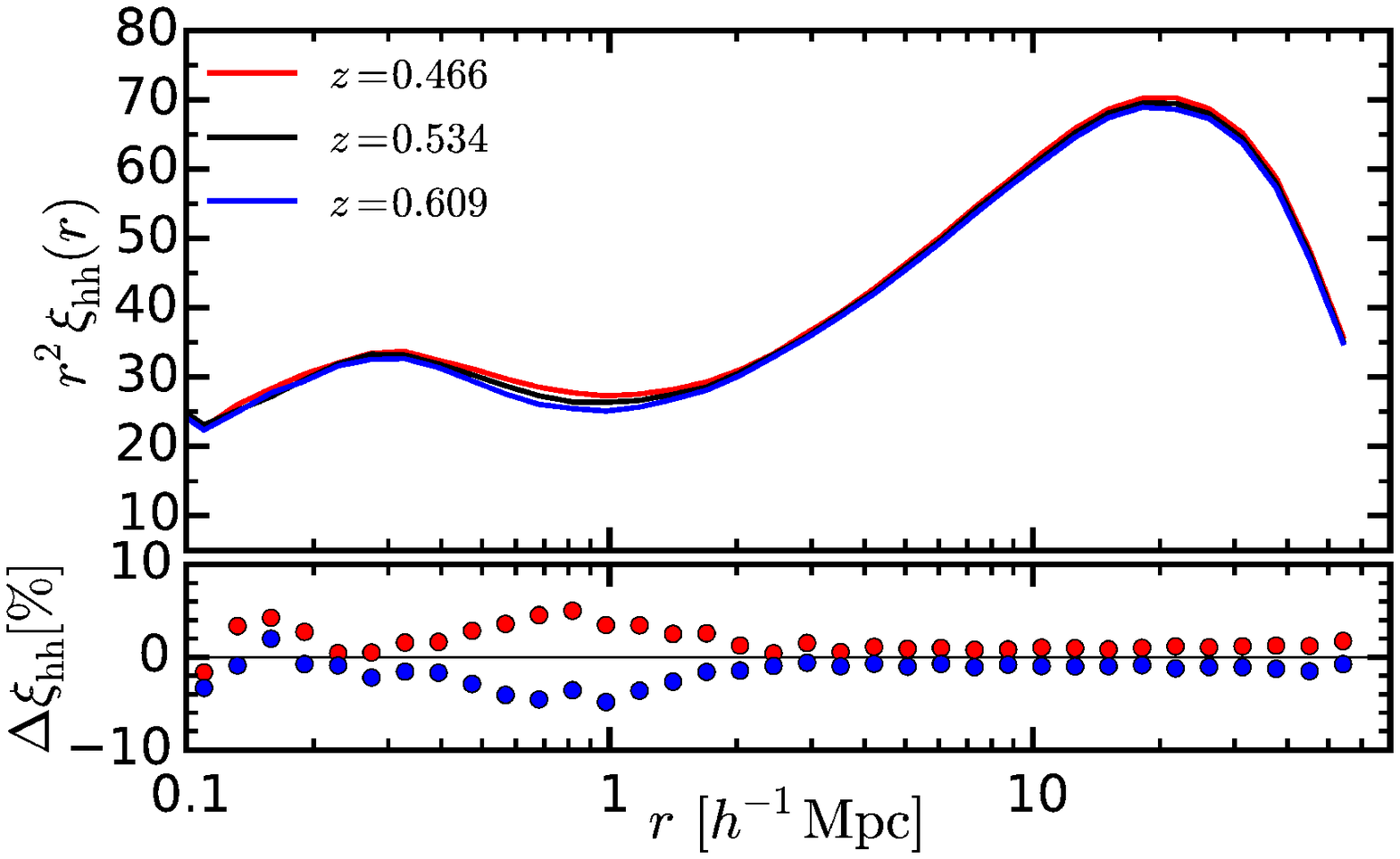}
\caption{\label{fig:Xi_hh_evolution} Time evolution of the clustering
  of subhalos at fixed number density. Subhalos are inversely sorted
  by $V_{\rm peak}$ and a cut is imposed at a number density of
  $\overline{n} \simeq 1.58\times 10^{-4} (h/{\rm Mpc})^{3}$. The
  upper panel shows a comparison of the three dimensional correlation
  function of subhalos in real space at different redshift outputs;
  $z=0.436$ (red), $z=0.534$ (default, black), and $z=0.609$
  (blue). The lower panel presents a fractional difference of the
  correlation function with respect to the one at the default
  $z=0.534$ output.}
\end{center}
\end{figure}

We begin by testing if a single redshift output is sufficient to model CMASS 
over the redshift range of $0.43<z<0.7$. 
We rank order subhalos by $V_{\rm peak}$ 
and select the top $N$ subhalos with a number density of 
$\overline{n} \simeq 1.58\times 10^{-4} (h^{-1}{\rm Mpc})^{-3}$. 
This value roughly corresponds to the number density of galaxies 
with $\log_{10}(M_{*}/M_{\odot})\gtrsim 11.0$.
Figure \ref{fig:Xi_hh_evolution} shows the three-dimensional 
correlation function of subhalos in real space 
as a function of separation at three different redshift outputs 
and at fixed number density $\overline{n}$. The correlation function 
varies by at most 5\% compared to $z=0.534$ over the CMASS redshift range.
The fractional difference at large scales, $r\gtrsim 3\,h^{-1}{\rm Mpc}$, 
is $1\operatorname{-}2$ \%. The largest differences (at the level of 5\%) 
are seen at the transition regime from the 2-halo to 1-halo term, 
$r\lesssim 1\,h^{-1}{\rm Mpc}$, where the errors on our observational clustering 
signal are increased by uncertainties due to the fiber-collision correction. 
In future work, especially when the $S/N$ of the measurements increase 
(currently we are using DR10 measurements), these effects will need to be taken into account. \par

\begin{figure}
\begin{center}
\includegraphics[width=1\columnwidth]{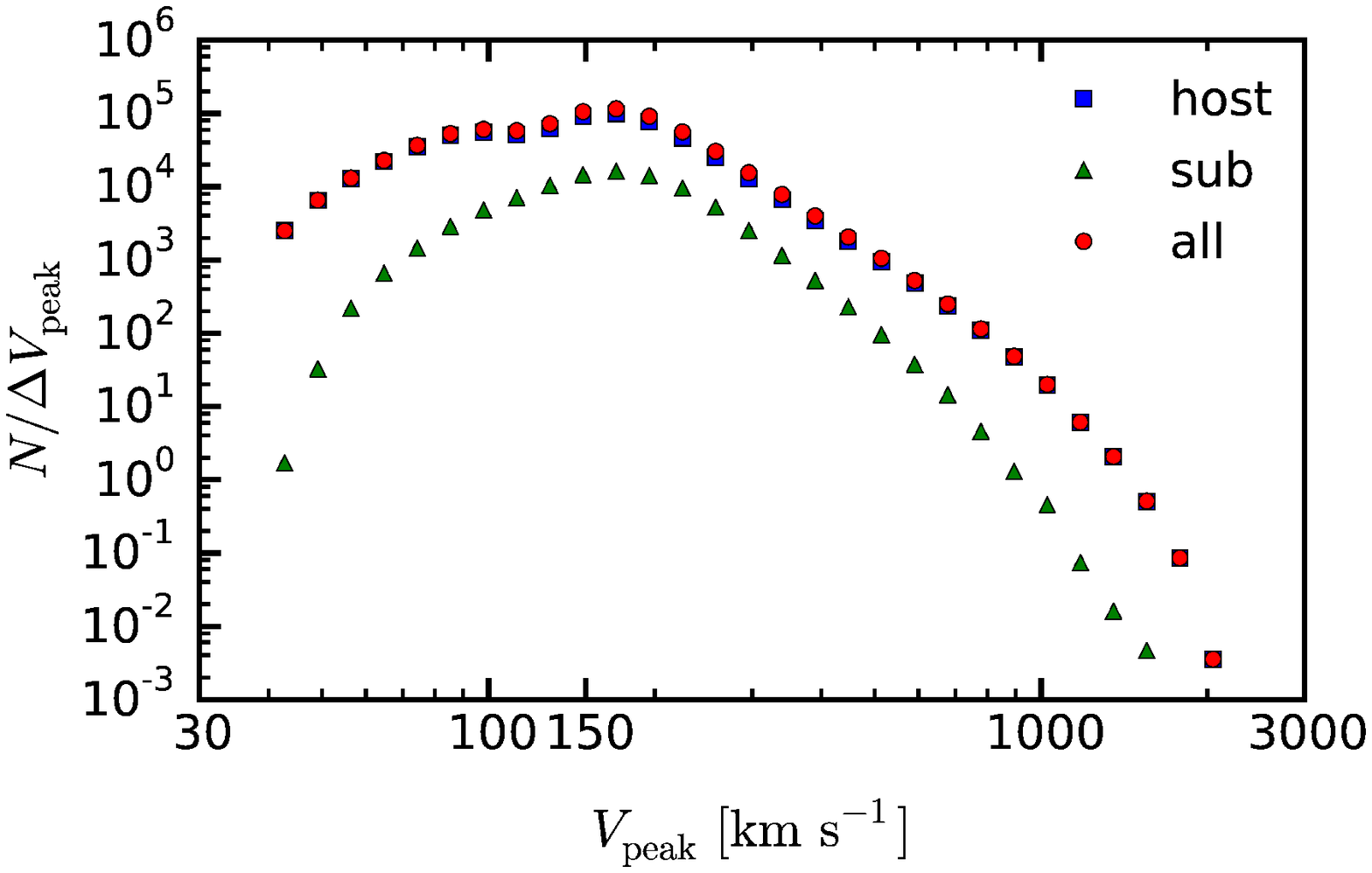}
\caption{\label{fig:resolution_Vpeak}
The histogram of host halos (blue square), subhalos (green triangle), 
and all halos (red circle) as a function of $V_{\rm peak}$. 
A clear turnover around $V_{\rm peak}\sim 150\,{\rm km\,s^{-1}}$ suggests 
that subhalos with $V_{\rm peak}\gtrsim 150\,{\rm km\,s^{-1}}$ are not affected 
by resolution. 
}
\end{center}
\end{figure}

We perform two tests concerning the impact of the resolution of {\sf MDR1} 
on our results. Based on \cite{White_2011} and also R14, 
we estimate that abundance matching for CMASS will require subhalos 
with $V_{\rm peak}\ge 200\,{\rm km\,s^{-1}}$. 
Figure \ref{fig:resolution_Vpeak} presents the histogram of subhalos 
as a function of $V_{\rm peak}$. This histogram starts to deviate from a power 
law at {\bf $V_{\rm peak}\sim 200\,{\rm km\,s^{-1}}$} and has a clear turnover 
at $V_{\rm peak}\sim 150\,{\rm km\,s^{-1}}$. Figure \ref{fig:resolution_Vpeak} demonstrates 
that {\sf MDR1} has a sufficient resolution for CMASS, 
although a higher resolution would be preferable.

\begin{figure}
\begin{center}
\includegraphics[width=1\columnwidth]{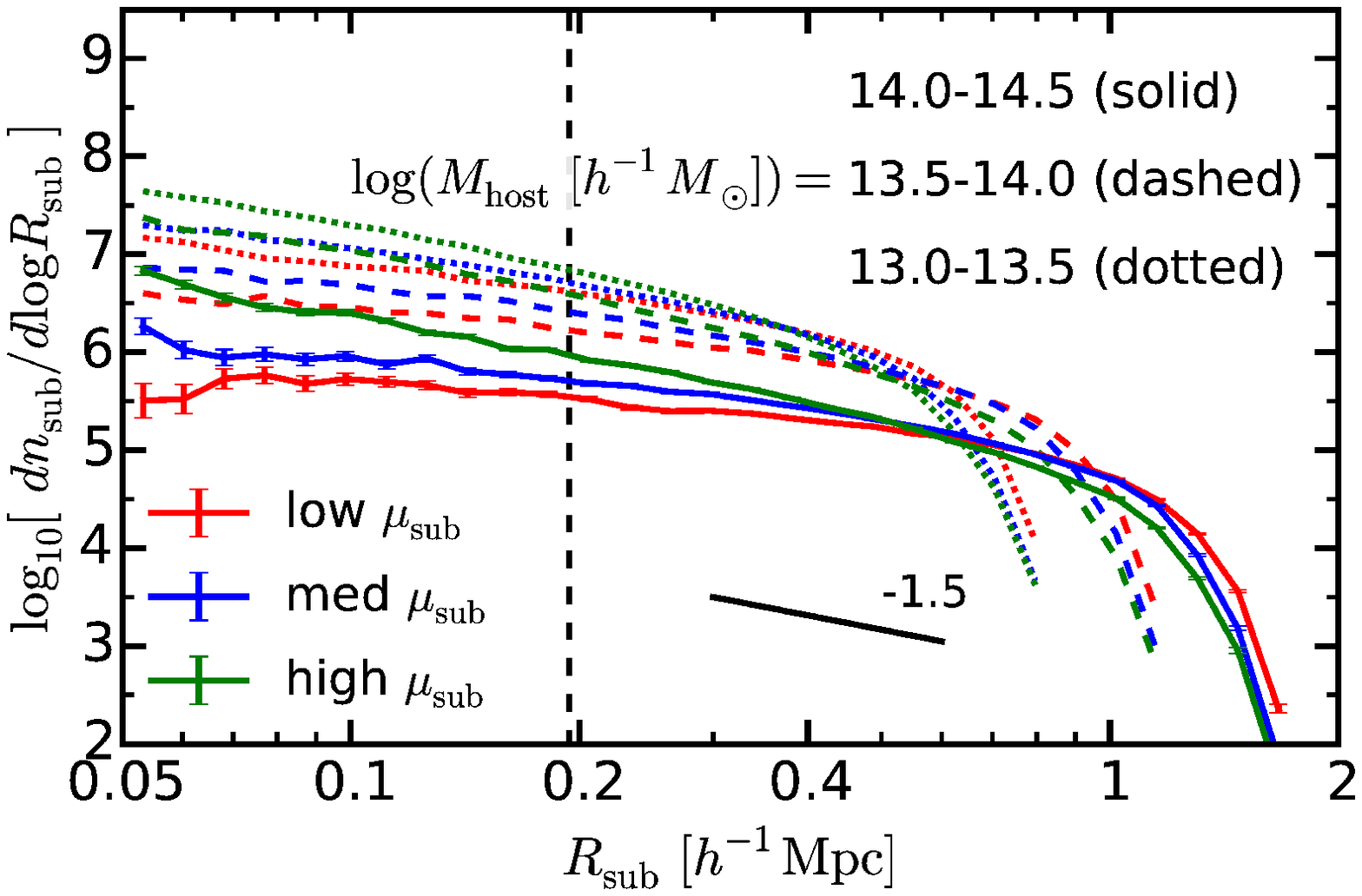}
\caption{\label{fig:resolution_Rsub}
Radial profiles of subhalos. Different colors correspond to different values of 
$\mu_{\rm sub}\equiv V^{\rm sub}_{\rm peak}/V^{\rm host}_{\rm peak}$.
The samples in terms of $\mu_{\rm sub}$ are created to contain an equal number of subhalos.
%divided such that $\mu_{\rm sub}$ contains the equal number. 
The {\it solid}, {\it dashed}, and {\it dotted } lines correspond to three 
different bins in host halo mass (but divided in terms of $V_{\rm peak}$). 
As a reference, the logarithmic slope of $-1.5$ is also shown. 
The radius at which subhalo detections are incomplete is estimated as the radius 
where the logarithmic slope of the profile becomes larger than -1.5.
The vertical black dashed line shows the minimum scale in our 
clustering measurement. %
}
\end{center}
\end{figure}

\begin{figure}
\begin{center}
\includegraphics[width=1\columnwidth]{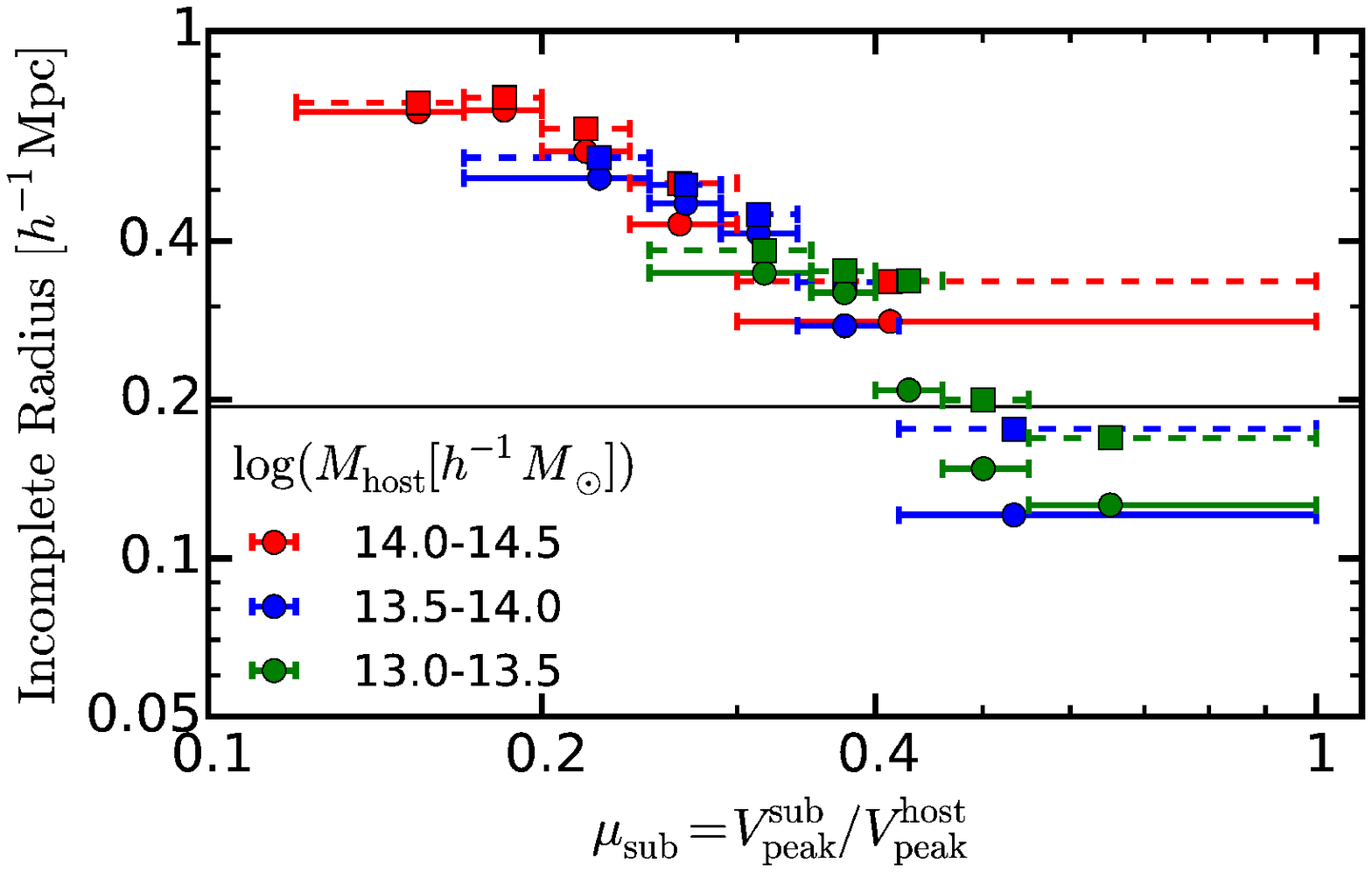}
\caption{\label{fig:Rsublimit}
Subhalo incompleteness radius as a function of $\mu_{\rm sub}$. Different colors indicate different bins in host halo mass. 
Circles with solid error bars show the results when the incomplete radius is defined with respect to a logarithmic slope of -1.5. 
Squares with dashed error bars represent the results when the incomplete radius is defined with respect to a logarithmic slope of -1.7. 
The horizontal black dashed line shows the minimum scale in our clustering measurement. 
Higher resolution simulations would be preferable and will be adopted in forthcoming paper.
}
\end{center}
\end{figure}

However, Figure \ref{fig:resolution_Vpeak} does not guarantee that the
resolution is sufficiently high to trust our clustering predictions down to
arbitrarily small scales. Our clustering signal is dominated by
central-satellite pairs in the 1-halo term regime, implying that it is
important to study the completeness of subhalos as a function of
distance to their host-hosts, $R_{\rm sub}$. Because the true radial
profiles of subhalos remain poorly known, it is difficult to
precisely characterize the radius at which incompleteness effects
become important. With this caveat in mind, \citet{Behroozi_2013a}
define the radius at which subhalo detections are incomplete as the
radius where the logarithmic slope of the profile becomes larger than
-1.5 (or -1.7). This cut-off is motivated by the density profiles of
observed subhalos in the maxBCG cluster catalog
\citep[][]{Tinker_2011}.  Figure \ref{fig:resolution_Rsub} displays the
radial profiles of subhalos for different ratios of $V_{\rm peak}$,
$\mu_{\rm sub}\equiv V^{\rm sub}_{\rm peak}/V^{\rm host}_{\rm peak}$,
and for three different bins in host halo mass (but divided by $V_{\rm
  peak}$).  In general, this radial profile becomes gradually
shallower at smaller $R_{\rm sub}$ due to the fact that density
contrast between the parent halo and subhalos decreases in the inner
regions of halos and subhalos become more difficult to detect. Using
the \citet{Behroozi_2013a} criterion, we estimate that subhalo
detections become incomplete at $0.1$-$0.7\,h^{-1}{\rm Mpc}$, depending on
$\mu_{\rm sub}$ and $M_{\rm host}$, as shown in Figure
\ref{fig:Rsublimit}.  The smallest scale in our $w_{p}$
measurement is $\approx 0.2\,h^{-1}{\rm Mpc}$ and is indeed close to the
incompleteness limit. We can definitely improve this situation by
using higher resolution simulations. However, we expect that the
impact of the resolution on our results should be relatively small,
since the errors of our measured $w_{p}$ on these scales are boosted
by systematic uncertainties in the fiber collision correction. We
conclude that the resolution of {\sf MDR1} is sufficient for our
purpose, but higher resolution simulations would be preferable
and will be adopted in subsequent work.

%=======================================================================%

%=======================================================================%

%%%%%%%%%%%%%%%%%%%%%%%%%%%%%%%%%%%%%%%%%%%%%%%%%%%%%%%%%%%%%%%%%%%%%%%%%%%%%%%%%%
%%%%%%%%%%%%%%%%%%%%%%%%%%%%%%%%%%%%%%%%%%%%%%%%%%%%%%%%%%%%%%%%%%%%%%%%%%%%%%%%%%
\section{Impact of the scatter in the $V_{\rm peak}{\mathchar`-}M_{*}$
  relation for the age-matching model}
\label{app:scatter_and_agm}
%%%%%%%%%%%%%%%%%%%%%%%%%%%%%%%%%%%%%%%%%%%%%%%%%%%%%%%%%%%%%%%%%%%%%%%%%%%%%%%%%%
%%%%%%%%%%%%%%%%%%%%%%%%%%%%%%%%%%%%%%%%%%%%%%%%%%%%%%%%%%%%%%%%%%%%%%%%%%%%%%%%%%

To study the impact of the assembly bias effect, we adopt the age-matching model 
where we reorganize the relation between subhalo age and galaxy color at fixed stellar mass 
rather than at fixed halo mass (or $V_{\rm peak}$). 
This is because we can perform rank-order matching only against
observable quantities. However, because our study operates in the very steep
end of the stellar mass function, we must verify that our results do
not depend on the stellar mass bin width when performing age-matching. 
In H13, the authors report that their analysis is insensitive to a stellar-mass bin width of 
$\Delta\log M_{*}=0.05{\mathchar`-}0.2$.
In this appendix, we perform a similar exercise to H13 for our extreme
age-matching model (see section \ref{subsec:AgMex}). In the following,
we demonstrate that our results are insensitive to our fiducial bin
width of $\Delta\log M_{*}=0.05$. However, we also show that the
choice of a fiducial bin width needs to take into consideration the
scatter (in our case $\sigma=0.105$). 

\begin{figure}
\begin{center}
\includegraphics[width=1\columnwidth]{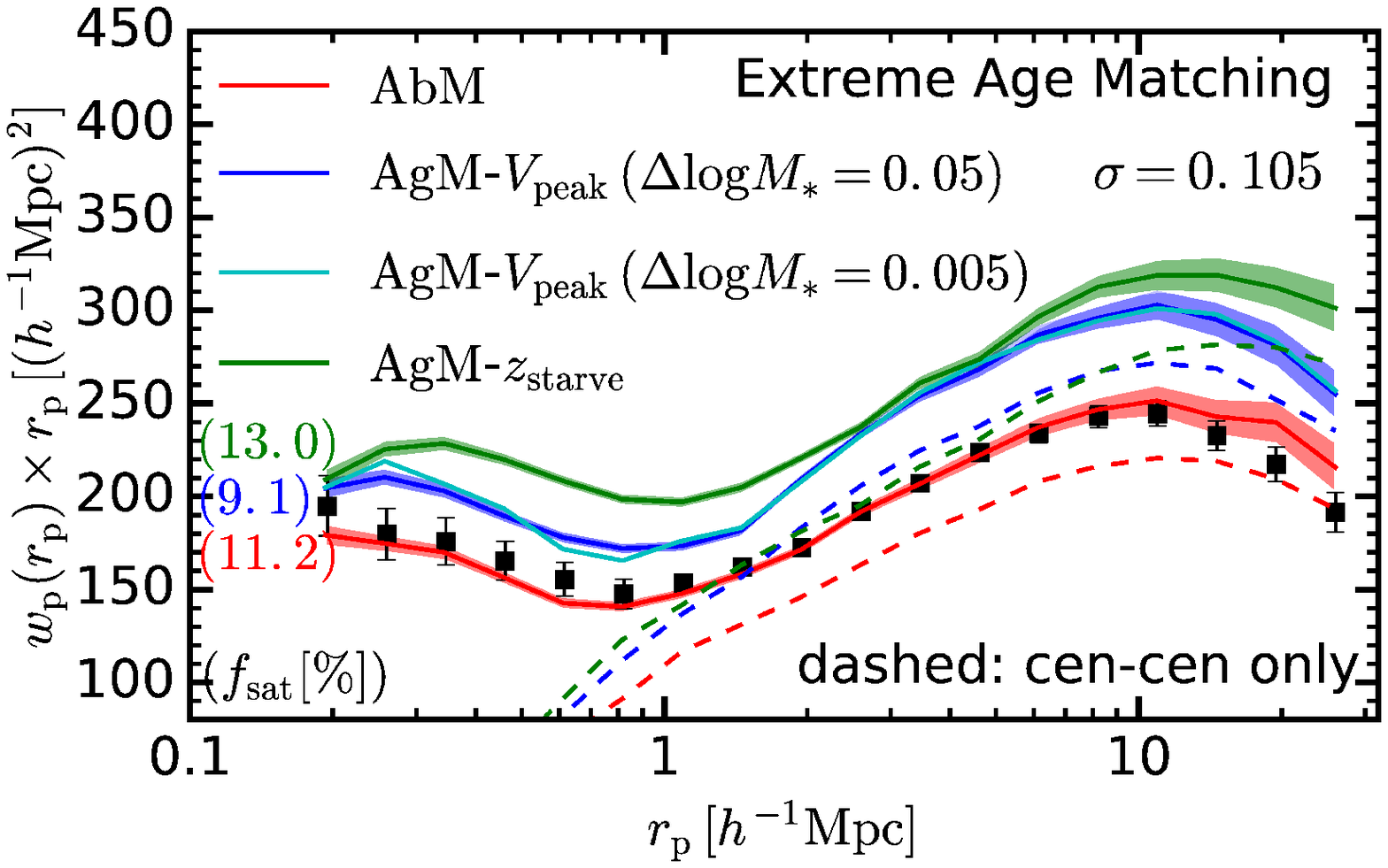}
\caption{\label{fig:test_ExAgM_Vpeak}
Testing our stellar-mass bin width in performing the age-matching model. 
We perform the age-matching model for the extreme case as discussed 
in section \ref{subsec:AgMex} but in terms of $V_{\rm peak}$ itself as a 
halo-age proxy. Note that the best-fitting values in the simple age matching, 
$(\phi_{1},\log_{10}M_{0},\sigma)=(1.86\times 10^{-3},10.89,0.105)$, 
are adopted here. 
$w_{\rm p}$ with the different bin size are shown in blue 
for $\Delta\log M_{*}=0.05$ and in cyan for 0.005, respectively. 
These results can be compared with the age-matching one (red) 
where a clear discrepancy with blue or cyan curve is confirmed. 
The age-matching model with $z_{\rm starve}$ is also shown just for 
a comparison with figure \ref{fig:AgM_wp}. 
}
\end{center}
\end{figure}

We perform a test in which we consider the extreme age-matching model 
in section \ref{subsec:AgMex} but we reshuffle with respect to $V_{\rm peak}$ 
rather than $z_{\rm starve}$. 
In addition, we test how the results vary if we use a different bin width. 
Figure \ref{fig:test_ExAgM_Vpeak} demonstrates that our results are insensitive 
to this change in bin width ($\Delta\log M_{*}=0.05$ (blue) and 
$\Delta\log M_{*}=0.005$ (cyan)). We have also checked that our mean halo masses 
and satellite fractions are unchanged when going from $\Delta\log M_{*}=0.05$ 
to $\Delta\log M_{*}=0.005$. 

Nevertheless, figure \ref{fig:test_ExAgM_Vpeak} shows a clear difference 
between the simple abundance matching (`AbM', red) and the extreme age matching results 
(`AgM-$V_{\rm peak}$', blue or cyan). 
In fact, the mean halo mass and the satellite fraction for the AbM (AgM-$V_{\rm peak}$) models are 
$\log (\overline{M_{\rm vir}}\,[M_{\odot}h^{-1}])=13.442$ (13.551), and $f_{\rm sat}=11.08\%$ (9.12\%), respectively. 
We argue that this difference originates from the non-zero scatter 
in the $V_{\rm peak}{\mathchar`-}M_{*}$ relation in the abundance matching. 
In performing the extreme age-matching model with $V_{\rm peak}$, CMASS galaxies with larger $X_{\rm col}$ 
at fixed stellar mass are likely to have larger $V_{\rm peak}$.

\begin{figure}
\begin{center}
\includegraphics[width=1\columnwidth]{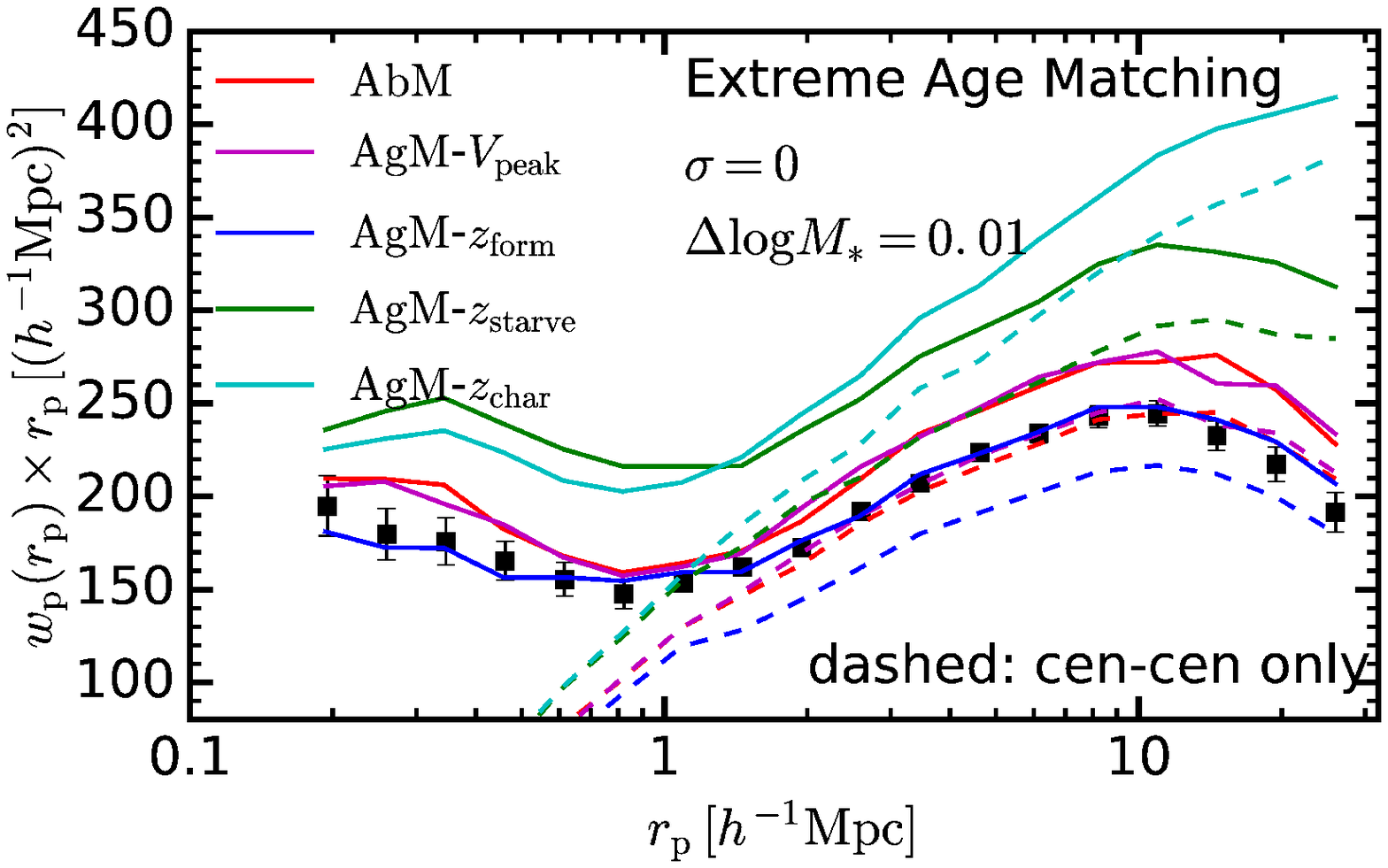}
\caption{\label{fig:test_ExAgM_Vpeak_sigma0}
Testing the impact of the scatter in the $V_{\rm peak}{\mathchar`-}M_{*}$ relation 
on the age-matching model. Here we fix the bin width with $\Delta\log M_{*}=0.01$ 
and do not introduce the scatter, i.e., $\sigma=0$. 
In this case, the abundance matching (red) and the age matching with $V_{\rm peak}$ 
(magenta) result in identical clustering. As a comparison, the age-matching results with 
$z_{\rm form}$ (blue), $z_{\rm starve}$ (green), and $z_{\rm char}$ (cyan) are also 
plotted to manifest the pure assembly bias effect in absence of the scatter. 
}
\end{center}
\end{figure}

Our argument is confirmed by figure \ref{fig:test_ExAgM_Vpeak_sigma0}
where we perform the same exercise but we set $\sigma=0$. In this
case, we find that our clustering prediction becomes stable with bin widths
smaller than $\Delta\log M_{*}=0.01$, and that AbM result is similar
to the AgM-$V_{\rm peak}$ one (compare red with magenta lines). In
figure \ref{fig:test_ExAgM_Vpeak_sigma0}, we also display the AgM
model with a variety of halo-age indicators (see section \ref{subsec:AgMex}). 
The halos masses of the blue, green and cyan curves are very similar 
($\log (\overline{M_{\rm vir}}\,[M_{\odot}h^{-1}])=13.513$). Hence, differences in the
clustering for the blue, green and cyan curves are a consequence of
assembly bias effects.

%=======================================================================%

%=======================================================================%

\bibliographystyle{mnras}
\bibliography{ms.bib}

\end{document}